\documentclass[3p]{elsarticle}

\usepackage{graphicx}
\usepackage{balance}  

\usepackage{hyperref}
\usepackage[lined,boxed,linesnumbered,commentsnumbered]{algorithm2e}

\usepackage{etoolbox}
\usepackage{comment}

\def\equationautorefname~#1\null{%
  (#1)\null
}

\usepackage{enumitem}
\usepackage{subcaption}
\usepackage[group-separator={,},group-minimum-digits=4]{siunitx}
\usepackage{multirow}
\usepackage{bbm}
\usepackage{diagbox}
\usepackage{xcolor}
\usepackage{array}
\usepackage{colortbl}
\usepackage{mathtools}
\usepackage{makecell}
\usepackage{csquotes}
\usepackage{amsthm}
\usepackage[nameinlink]{cleveref}
\usepackage{chngcntr}
\usepackage{amssymb}
\usepackage{booktabs}

\usepackage{tikz-cd}

\newcommand{\RN}[1]{%

\textup{\uppercase\expandafter{\romannumeral#1}}}

\usetikzlibrary{decorations.pathmorphing}

\newcommand{\appendixref}[1]{\autoref{#1}}

\newcommand{\RNum}[1]{\uppercase\expandafter{\romannumeral #1\relax}}
\newcommand{\Rnum}[1]{\lowercase\expandafter{\romannumeral #1\relax}}

\usepackage{etoolbox}
\newtheoremstyle{saber}
  {}
  {}
  {\itshape}
  {}
  {\bfseries}
  {.}
  {.5em}
  {}

\theoremstyle{saber}
\ifcsundef{theorem}{
\newtheorem{theorem}{Theorem}
}{}
\ifcsundef{corollary}{%
}{}
\ifcsundef{lemma}{%
}{}
\ifcsundef{definition}{%
\newtheorem{definition}[theorem]{Definition}
}{}
\ifcsundef{fact}{%
\newtheorem{fact}[theorem]{Fact}
}{}
\ifcsundef{remark}{%
}{}
\ifcsundef{property}{%
}{}
\ifcsundef{assumption}{%
\newtheorem{assumption}[theorem]{Observation}
}{}
\ifcsundef{example}{%
\newtheorem{example}[theorem]{Example}
}{}
\ifcsundef{modification}{%
\newtheorem{modification}[theorem]{Modification}
}{}

\ifx\theoremautorefname\undefined
\newcommand*{\theoremautorefname}{Theorem}
\fi
\ifx\lemmaautorefname\undefined
\newcommand*{\lemmaautorefname}{Lemma}
\fi
\ifx\definitionautorefname\undefined
\newcommand*{\definitionautorefname}{Definition}
\fi
\ifx\corollaryautorefname\undefined
\newcommand*{\corollaryautorefname}{Corollary}
\fi
\ifx\factautorefname\undefined
\newcommand*{\factautorefname}{Fact}
\fi
\ifx\propertyautorefname\undefined
\newcommand*{\propertyautorefname}{Property}
\fi

\ifx\argmax\undefined
\newcommand{\argmax}{\operatornamewithlimits{argmax}}
\fi
\ifx\argmin\undefined
\newcommand{\argmin}{\operatornamewithlimits{argmin}}
\fi
\ifx\limsup\undefined
\newcommand{\limsup}{\operatornamewithlimits{limsup}}
\fi
\ifx\liminf\undefined
\newcommand{\liminf}{\operatornamewithlimits{liminf}}
\fi
\ifx\norm\undefined
\newcommand{\norm}[1]{\lVert#1\rVert}
\fi
\ifx\abs\undefined
\newcommand{\abs}[1]{\lvert#1\rvert}
\fi
\ifx\set\undefined
\newcommand{\set}[1]{\left\{#1\right\}}
\fi
\ifx\mset\undefined
\newcommand{\mset}[1]{\lbrack #1\rbrack}
\fi
\ifx\etal\undefined
\newcommand{\etal}[1]{{\em #1 et al.}~}
\fi
\ifx\ie\undefined
\newcommand{\ie}{{\em i.e.,} }
\fi
\ifx\eg\undefined
\newcommand{\eg}{{\em e.g.,} }
\fi
\ifx\etc\undefined
\newcommand{\etc}{{\em etc.,} }
\fi
\ifx\wrt\undefined
\newcommand{\wrt}{{\em w.r.t.} }
\fi
\ifx\dotprod\undefined
\newcommand{\dotprod}[2]{
  \langle #1, #2 \rangle
}
\fi
\ifx\iid\undefined 
\newcommand{\iid}{i.i.d.}
\fi
\ifx\bigParenthes\undefined 
\newcommand{\bigParenthes}[1]{
  \big(#1\big)
}
\fi
\ifx\bigBracket\undefined 
\newcommand{\bigBracket}[1]{
  \big\{#1\big\}
}
\fi
\ifx\bigSqBracket\undefined 
\newcommand{\bigSqBracket}[1]{
  \big[#1\big]
}
\fi
\ifx\BigParenthes\undefined 
\newcommand{\BigParenthes}[1]{
  \Big(#1\Big)
}
\fi
\ifx\BigBracket\undefined 
\newcommand{\BigBracket}[1]{
  \Big\{#1\Big\}
}
\fi
\ifx\BigSqBracket\undefined 
\newcommand{\BigSqBracket}[1]{
  \Big[#1\Big]
}
\fi
\ifx\biggParenthes\undefined 
\newcommand{\biggParenthes}[1]{
  \bigg(#1\bigg)
}
\fi
\ifx\biggBracket\undefined 
\newcommand{\biggBracket}[1]{
  \bigg\{#1\bigg\}
}
\fi
\ifx\biggSqBracket\undefined 
\newcommand{\biggSqBracket}[1]{
  \bigg[#1\bigg]
}
\fi
\ifx\BiggParenthes\undefined 
\newcommand{\BiggParenthes}[1]{
  \Bigg(#1\Bigg)
}
\fi
\ifx\BiggBracket\undefined 
\newcommand{\BiggBracket}[1]{
  \Bigg\{#1\Bigg\}
}
\fi
\ifx\BiggSqBracket\undefined 
\newcommand{\BiggSqBracket}[1]{
  \Bigg[#1\Bigg]
}
\fi
\ifx\bracket\undefined
\newcommand{\bracket}[1]{
  \{#1\}
}
\fi
\ifx\parenthes\undefined
\newcommand{\parenthes}[1]{
  (#1)
}
\fi
\ifx\sqBracket\undefined
\newcommand{\sqBracket}[1]{
  [#1]
}
\fi
\ifx\prob\undefined 
\newcommand{\prob}[1]{\mathbb{P}[#1]}
\fi
\ifx\Prob\undefined 
\newcommand{\Prob}[1]{\mathbb{P}\big[#1\big]}
\fi
\ifx\probb\undefined 

\fi
\ifx\expect\undefined 
\newcommand{\expect}[1]{\mathbb{E}[#1]}
\fi
\ifx\Expect\undefined 
\newcommand{\Expect}[1]{\mathbb{E}\big[#1\big]}
\fi
\ifx\expectt\undefined 
\newcommand{\expectt}[1]{\mathbb{E}\bigg[#1\bigg]}
\fi
\ifx\walk\undefined 
\makeatletter
\newcommand{\walk}[1]{%
  \@tempswafalse
  \@for\next:=#1\do
    {\if@tempswa\!\!\rightarrow\!\!\else\@tempswatrue\fi\next}%
}
\makeatother
\fi
\ifx\seq\undefined 
\newcommand{\seq}{\!=\!}
\fi
\ifx\sminus\undefined 
\newcommand{\sminus}{\!-\!}
\fi
\ifx\sm\undefined 
\newcommand{\sm}[1]{\!#1\!}
\fi

\ifx\union\undefined
\newcommand{\union}[2]{#1\!\cup\!#2}
\fi

\def\solutionname{\mbox{\textsc{CommonSense}}\xspace}
\def\thistitle{\solutionname{}: Efficient Set Intersection (SetX) Protocol Based on Compressed Sensing}
\def\problemname{SetX}

\def\problemnamelong{set intersection}
\def\reconnamelong{set reconciliation}

\def\reconname{SetR}
\def\indicator{\mathbf{1}}
\def\alisset{\mathsf{A}}
\def\alisvec{\indicator_\alisset}
\def\alissize{|\alisset|}
\def\belaset{\mathsf{B}}
\def\belavec{\indicator_\belaset}
\def\belasize{|\belaset|}
\def\alismbela{{\alisset \setminus \belaset}}
\def\belamalis{{\belaset \setminus \alisset}}
\def\abunion{{\alisset \cup \belaset}}
\def\absymdiff{{\alisset \triangle \belaset}}
\def\hbelamalis{{\widehat\belamalis}}
\def\halismbela{{\widehat\alismbela}}
\def\abintersec{{\alisset \cap \belaset}}
\def\coirset{\mathsf{C}}
\def\sensemat{\mathbf{M}}
\def\sensecol{\vec{\mathbf{m}}}
\def\diffsize{d}

\def\basevec{\vec{\mathbf{e}}}
\def\noise{\vec{\mathbf{\epsilon}}}

\def\counternum{l}
\def\fanout{m}
\def\numcols{n}
\def\cscode{\vec{\mathbf{r}}}

\def\aset{\mathsf{S}}
\def\bitwidth{u}
\def\universe{\mathsf{U}}

\def\vector{\vec{\mathbf{v}}}

\def\sigcor{x}
\def\signal{\vec{\mathbf{x}}}

\begin{document}

\title{\thistitle{}}

\author[1]{Jingfan Meng\corref{corresponding}}
\cortext[corresponding]{Corresponding Authors}
\ead{jmeng40@gatech.edu}
\author[1]{Tianji Yang\corref{corresponding}}
\ead{tyang425@gatech.edu}
\author[1]{Jun Xu\corref{corresponding}}
\ead{jx9@gatech.edu}

\address[1]{School of Computer Science, Georgia Institute of Technology,
  Atlanta, USA}
\begin{abstract}
  Set reconciliation (SetR) is an important research problem that has been
  studied for over two decades.
  In this problem, two large sets $\mathsf{A}$ and $\mathsf{B}$ of objects
  (tokens, files, records, etc.)
  are stored respectively at two different network-connected hosts, which we name
  Alice and Bob respectively.
  Alice and Bob need to communicate with each other to learn the set union
  $\mathsf{A} \cup \mathsf{B}$ (which then becomes their reconciled state), at
  low communication and computation costs.
  In this work, we study a different problem intricately related to SetR: Alice
  and Bob collaboratively compute $\mathsf{A} \cap \mathsf{B}$.
  We call this problem SetX (set intersection).
  Although SetX is just as important as SetR, it has never been properly studied
  in its own right.
  Rather, there is an unspoken perception by the research community that SetR and
  SetX are equally difficult (in costs), and hence ``roughly equivalent.''
  Our first contribution is to show that SetX is fundamentally a much ``cheaper''
  problem than SetR, debunking this long-standing perception.
  Our second contribution is to develop a novel SetX solution, the
  communication cost of which handily beats the information-theoretic lower bound
  of SetR.
  This protocol is based on the idea of compressed sensing (CS), which we
  describe here only for the special case of $\mathsf{A} \subseteq \mathsf{B}$
  (We do have a more sophisticated protocol for the general case).
  Our protocol is for Alice to encode $\mathsf{A}$ into a CS sketch $\mathbf{M}
    \mathbf{1}_{\mathsf{A}}$ and send it to Bob, where $\mathbf{M}$ is a CS matrix
  with $l$ rows and $\mathbf{1}_{\mathsf{A}}$ is the binary vector representation
  of $\mathsf{A}$.
  Our key innovation here is to make $l$ (the sketch size) just large enough (for
  the sketch) to summarize $\mathsf{B} \setminus \mathsf{A}$ (what Alice misses).
  In contrast, in existing protocols $l$ needs to be large enough to summarize
  $\mathsf{A}$ (what Alice knows), which is typically much larger in cardinality.
  Our third contribution is to design a CS matrix $\mathbf{M}$ that is both
  ``friendly'' to (the performance of) applications and ``compliant'' with CS
  theory.
\end{abstract}
\maketitle

\section{Introduction}\label{sec:intro}

Set reconciliation (\reconname{}) is a fundamental algorithmic problem that
arises in many networking, system, and database applications, and hence has
been studied for over two
decades~\cite{Minsky2003,Dodis_PinSketch_2008,Guo_SetReconciliationvia_2013,
  Luo_SetReconciliationCuckoo_2019,Eppstein_WhatsDifference_2011, gong-pbs}.
In this problem, two large sets $\alisset$ and $\belaset$ of objects (tokens,
files, records, etc.) are stored respectively at two different
network-connected hosts, which we name Alice and Bob respectively.
Alice and Bob need to communicate with each other to learn the set union
$\abunion$, which then becomes their reconciled state, and to do so at low
communication (between Alice and Bob) and computation (at both Alice and Bob)
costs.

In this work, we study a different research problem that is intricately related
to SetR.
In this problem, which we call \problemname{} (\emph{\problemnamelong{}}),
Alice and Bob need to collaboratively compute $\abintersec$ (instead of
$\abunion$).
One connection between \reconname{} and \problemname{} is the following fact:
None of the \emph{existing} \reconname{} solutions (which we will review
in~\autoref{ssec:related-recon}) can let one host, say Bob, compute the union
(to which end Bob needs to know Alice's \emph{unique elements} $\alismbela$)
without also letting him know the intersection (or equivalently, revealing his
own unique elements $\belamalis$).

\subsection{The First Contribution: \problemname{}
  Is Fundamentally Cheaper Than \reconname{}}\label{sec:psea} Perhaps due to the
above fact, there is an unspoken perception by the research community that the
two research problems \reconname{} and \problemname{} are equally difficult and
expensive (in communication and computation costs).
Indeed, this perception is in full display in a fairly recent \reconname{}
solution called Graphene~\cite{Ozisik2019}.
In Graphene, Alice and Bob need to solve \problemname{} in the special case of
$\alisset\subseteq \belaset$.
The proposed solution, to be described in~\autoref{ssec:setx-related}, employs a
\emph{full-fledged} \reconname{} protocol as a subroutine, even though it only
uses the ``\problemname{} part'' of this subroutine's output.
Had the Graphene authors questioned this perception, they would have searched
for a much cheaper \problemname{}-specific subroutine as we do in this paper.

For the reasons above, it is fair to say that \problemname{} has never been
properly studied in its own right, although SetX has just as many applications
as SetR that we will elaborate in~\autoref{sec:motivation}.
The \emph{first contribution} of this paper is to show conclusively that this
long-standing perception is not correct, and that \problemname{} is
fundamentally a much ``cheaper'' problem than \reconname{}.
We do so by establishing a large gap between the information-theoretic lower
bounds on the communication costs of \reconname{} and \problemname{} (as we
will show shortly).
Moreover, we show that a practical \problemname{} protocol (which will be our
next contribution) handily beats the information-theoretic lower bound of the
\reconname{} established in~\cite{Minsky2003}.


Before describing the lower bounds, we introduce some notations on set
cardinalities and state an assumption implicitly made in all existing studies
of
\reconname{}~\cite{Eppstein_WhatsDifference_2011,Dodis_PinSketch_2008,yang-rateless-iblt}.
We denote Alice's (set's) cardinality by $\alissize$, Bob's cardinality by
$\belasize$, and the \emph{symmetric difference cardinality} (SDC) by
$\diffsize \triangleq |\absymdiff|$, wherein $ \absymdiff \triangleq
  (\alismbela)\bigcup (\belamalis)$ is the symmetric difference between
$\alisset$ and $\belaset$.
The assumption is that $\diffsize \ll \alissize$ and $\diffsize \ll \belasize$.
In other words, $\alisset$ and $\belaset$ are much larger than their SDC.
This assumption holds in virtually all practical applications of \reconname{}.
For example, in file synchronization, the difference between two files is
usually much smaller than the two files per se~\cite{tridgell-rsync}.
We continue to use this assumption in studying \problemname{}, because it also
holds in virtually all applications of \problemname{}.

In fact, without this assumption, \reconname{} would not be an interesting
problem (that is worth studying), for the following reason.
When this assumption does not hold (or when $|\absymdiff|$ is on the same order
as $\alissize$ and $\belasize$), the communication cost of the trivial protocol
of sending $\alisset$ and $\belaset$ respectively (in its entirety or its
\emph{set membership filter} summary, see~\autoref{ssec:membership}) to each
other is already close to the information-theoretic lower bound of
\reconname{}, so all advanced protocols (despite having high computation costs)
can only achieve meager savings.
The same statement and reason can also be said about \problemname{}.

For the \problemname{} problem, we further assume that the size of the universe
$\universe$ is much larger than $\alissize$ and $\belasize$.
This assumption holds for virtually all applications of \problemname{} and
\reconname{} for the following reason: The identifier of an object is usually
its hash value (say $\bitwidth$ bits long, where $|\universe| = 2^\bitwidth$),
which needs to be long enough so that no two distinct objects in $\abunion$ can
have the same hash value (with a ``surprisingly'' high probability predicted by
the birthday paradox~\cite{flajo-birthday}).
For example, in the Ethereum dataset in~\autoref{ssec:ethereum}, the size of
the universe is $2^{256}$ (as the identifier of each Ethereum account is a
256-bit-long hash value), whereas each snapshot (of the world state) contains
only around 280 millions ($\approx 2^{28}$) distinct accounts.

Finally, we add a ``benign'' assumption for the ease of presentation:
$\alissize \le \belasize$ without loss of generality (since otherwise, we can
swap the roles of ``Alice'' and ``Bob'').

We now formalize these three assumptions into the following inequality.
\begin{equation}\label{eq:assumption}
  \diffsize \ll \alissize \le \belasize \ll 2^\bitwidth.
\end{equation}

As we will show in~\autoref{sec:with-set-recon}, the information-theoretic
lower bound on the communication cost of \problemname{} is roughly
$\diffsize\log_2 (e\alissize/\diffsize)$ bits ($e$ is the base of natural
logarithm), which is lower than that of \reconname{}, which is $\diffsize
  \log_2(e|\universe|/\diffsize)$ bits by~\cite{Minsky2003}, by a factor of 24.8
in the above Ethereum example (\SI{1.2}{MB} vs. {29.7}{MB}, assuming
$\diffsize$ is one million).
In fact, the communication cost of our proposed protocol also easily beats the
lower bound of \reconname{} (as we will show in \Cref{exam:uni} and
\Cref{exam:bi}).

Since this work is the first to show the surprising fact that \problemname{} is
fundamentally much cheaper, in terms of communication cost, than \reconname{},
our solution \solutionname{} is optimized primarily for communication cost, to
support this fact or narrative.
We emphasize, however, that \solutionname{} has paid only a modest price for
this primary optimization objective.
This price is the increased computation costs for (1) CS encoding and decoding;
and (2) entropy compression (using BCH~\cite{bch1} and asymmetric numerical
systems~\cite{duda-ans}) of the first message (from Alice to Bob).
The price appears to be modest: As will be shown in~\autoref{ssec:ethereum},
our \solutionname{} is only a couple of times slower than
D.Digest~\cite{Eppstein_WhatsDifference_2011} (described
in~\autoref{ssec:related-recon}), the fastest solution for \reconname{} (and
\problemname{} as its ``subsidiary'' problem).
In comparison, PinSketch (to be described in~\autoref{ssec:related-recon}), the
\reconname{} solution optimized primarily for communication cost, is already
slower than D.Digest by two orders of magnitude when $\diffsize > 1000$ (and
more so when $\diffsize$ is larger), as shown in Figure 1(d)
in~\cite{gong-pbs}.

\subsection{Second Contribution: \solutionname{}
  Protocol for \problemname{}}\label{ssec:second-contrib}

In
this section, we describe the second contribution of this work: a novel
solution to \problemname{}.
We call our solution \solutionname{} because of the following two facts.
First, in \problemname{} Alice and Bob learns their \emph{common} elements ($\abintersec$).
Second, our solution is based on the idea of compressed \emph{sensing}
(CS)~\cite{candes-rip}.

To describe \solutionname{}, we first provide a crash course on CS.
For reasons that will be clear later, we need to recover a
\emph{$\diffsize$-sparse} (containing $\diffsize$ nonzero scalars)
\emph{signal} vector $\signal$, which lies in a high-dimensional space
$\mathcal{S}$ (say of dimension $\numcols$), from an indirect \emph{linear
  measurement} (vector) $\cscode\triangleq \sensemat \signal$ of it, wherein
$\sensemat$ is an $\counternum\times \numcols$ \emph{CS matrix} (which is
randomly generated and fixed hereafter).
This vector $\cscode$ is often also called a {\it linear sketch}, because in
many CS applications it can be viewed as sketching (succinctly summarizing) a
sparse (system state) vector $\signal$ using a suitable linear transformation
$\sensemat$.
We introduce the term {\it sketch} here solely for the ease of presentation in
the sequel.
Although there is indeed a deep connection between CS and data
streaming/sketching~\cite{lu-counter-braids}, this connection is mostly
orthogonal to \solutionname{}.

A key result in CS theory is that, when $\sensemat$ satisfies a nice property
(called \emph{restricted isometry property}, or RIP, which we will elaborate in
\Cref{def:rip}), with high probability the dimension $\counternum$ of the
linear measurement $\cscode$ only needs to be slightly larger than $\diffsize$,
or more precisely $\counternum = O(\diffsize \log(\numcols/\diffsize))$, to
mathematically ``pin down'' $\signal$, in the sense $\signal$ is the unique
\emph{sparse} solution to the system of linear equations $\cscode= \sensemat
  \vector$ (here, the word ``sparse'' is essential, since by $\counternum\ll
  \numcols$, there are infinitely many dense solutions).
Furthermore, we can recover (compute) the correct sparse solution $\signal$
using MP (matching pursuit, to be elaborated in \autoref{algo:omp}), a family
of fast decoding algorithms.
This computation is called {\it sparse recovery} in the CS literature.

\begin{figure}
  \begin{subfigure}[b]{0.47\textwidth} \centering
    \includegraphics[width=\textwidth]{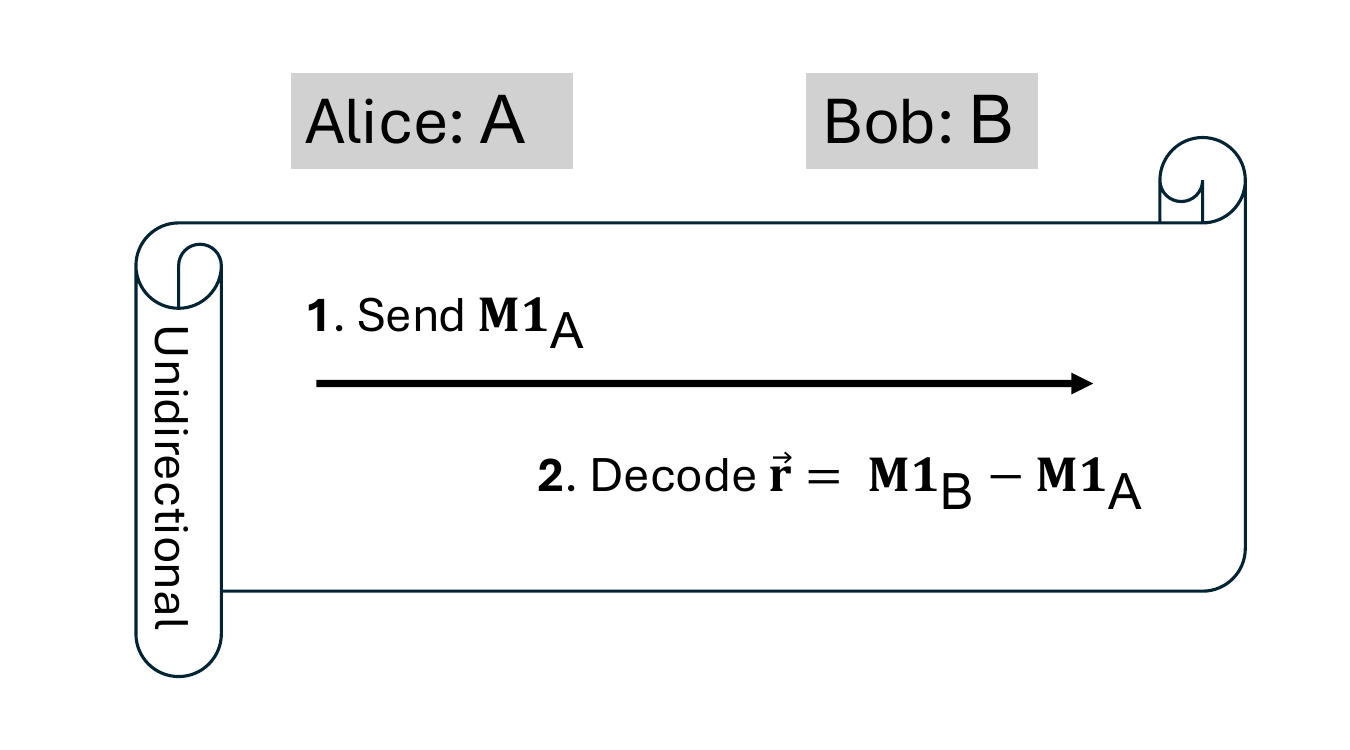}
    \caption{Base protocol for unidirectional \problemname{}
      ($\alisset \subseteq \belaset$,~\autoref{sec:special-model}).
    }\label{fig:uni-work}
  \end{subfigure}
  \begin{subfigure}[b]{0.47\textwidth}
    \centering
    \includegraphics[width=\textwidth]{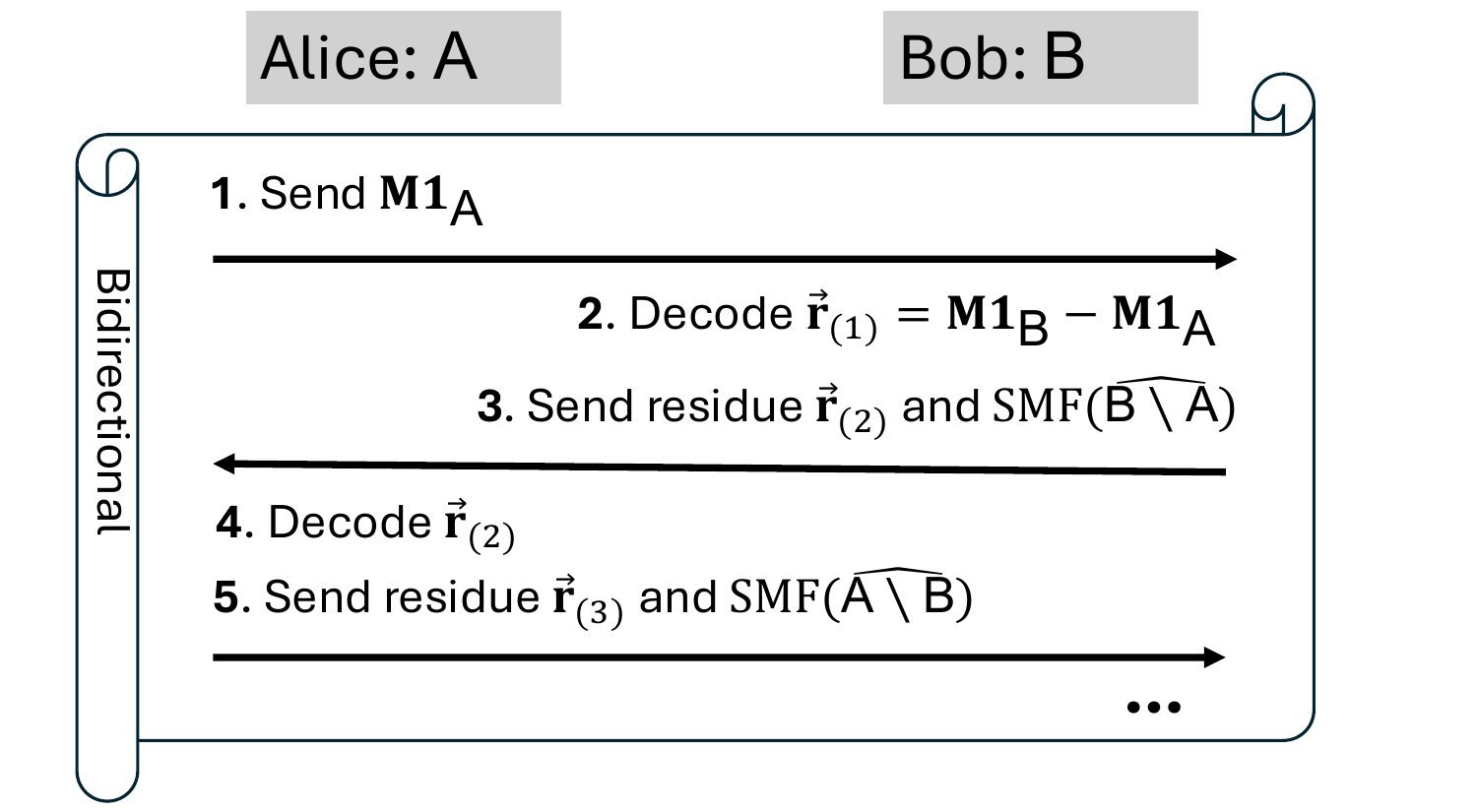}
    \caption{
      Extended protocol for bidirectional \problemname{}
      ($\alisset \not\subseteq \belaset$,~\autoref{sec:model}).
    }\label{fig:bi-work}
  \end{subfigure}
  \caption{Workflow of \solutionname{}.
    SMF refers to a set membership filter (see~\autoref{ssec:membership}), for
    example a Bloom filter.
  }\label{fig:workflow}
\end{figure}

Here, we describe a strawman CS-based \solutionname{} protocol only for the
special (\emph{unidirectional}) case of $\alisset \subseteq \belaset$ while
postponing the presentation of the \solutionname{} protocol for the general
(\emph{bidirectional}) case of $\alisset\not\subseteq \belaset$ and $\belaset
  \not\subseteq \alisset$ until~\autoref{sec:model}.
Our protocol (\Cref{fig:uni-work}) is for Alice to encode $\alisset$ into a CS
sketch vector $\sensemat \alisvec$ (an $\counternum$-dimensional linear
measurement) in step \textbf{1}, where $\alisvec$ denotes an
$|\alisset|$-sparse binary vector representation of $\alisset$ (to be precisely
defined in~\autoref{ssec:special-overview}), and so on for
$\indicator_\belamalis$ and $\belavec$ that will appear shortly.
Once Bob receives $\sensemat \alisvec$, he computes $\sensemat
  \indicator_\belamalis$ as $\sensemat \belavec$ (which Bob knows) minus
$\sensemat \alisvec$ (step \textbf{2}).

We need to highlight a fact here: Since the signal that Bob tries to recover is
$\indicator_\belamalis$, $\counternum$ only needs to be slightly larger than
$|\belamalis|$, for him to losslessly recover $\belamalis$ with high
probability empirically (so that he can compute $\abintersec =
  \belaset\setminus (\belamalis)$).
In other words, in our solution, $\counternum$ only needs to be large enough
(for the sketch) to summarize $\belamalis$ (what Alice misses).
In contrast, in existing \problemname{} protocols such as Graphene,
$\counternum$ (or its equivalence) needs to be large enough to summarize
$\alisset$ (what Alice knows), which is typically much larger in cardinality as
we have assumed in~\autoref{sec:psea}.
Thanks to this difference (that significantly reduces $\counternum$) and an
entropy compression scheme (to be described in\appendixref{sec:compress}) that
encodes each of the $\counternum$ scalars in $\sensemat \alisvec$ using as few
bits as possible, our \solutionname{} protocol is extremely
communication-efficient and can handily beat the information-theoretic lower
bound of the SetR problem, as will be shown
in~\autoref{sssec:bidirectional-eval}.

\subsection{Third Contribution: Friendly and Compliant CS Matrix}

In our third contribution, we identify a CS matrix $\mathbf{M}$ that
simultaneously has the following two desirable properties.
First, $\mathbf{M}$ needs to be ``friendly'' to (the performance of) the
networking applications of CS in general, and our \solutionname{} protocol in
particular in the following sense:  The computation cost of updating 
the compressed-sensed signal $\mathbf{M}\signal$ for a {\it 1-sparse change} (to be defined shortly) 
should be very low.
Second, $\mathbf{M}$ needs to be ``compliant'' with CS theory in the sense
it needs to satisfy a certain form of the aforementioned RIP, so that
using $\mathbf{M}$ allows us to ``provably benefit'' from the correctness and
(decoding) time complexity guarantees established in the CS theory.
However, to satisfy both properties turns out to be a tall order because the
  ``friendly'' property generally demands that $\mathbf{M}$ be a binary sparse
  matrix whereas the ``compliant'' property demands that $\mathbf{M}$ be a dense
  matrix, as we will elaborate next.
We feel incredibly lucky to find an $\mathbf{M}$ that manages to ``walk the
tightrope'' between these two properties.

We first explain why being ``compliant'' demands $\mathbf{M}$ to be dense.
As mentioned earlier, for the sparse recovery to succeed (in computing the
correct $\belamalis$), $\sensemat$ needs to satisfy the RIP, which, by
\Cref{def:rip}, means that $\sensemat$ approximately preserves the $L_2$
distance when operating on (multiplying to) any vector $\vector$.
All CS matrices that satisfy RIP are dense~\cite{chandar-negative}, with the
Gaussian random matrix being a notable example.

However, to be ``friendly'' to networking applications, it is much more
desirable for $\sensemat$ to be sparse than to be dense, for the following two
reasons.
First, in some networking applications of CS, the signal $\signal$ to be
measured often needs to be constantly updated in a 1-sparse manner, that is,
$\signal_{new} \gets \signal_{old} + \eta \basevec_i$ where $\eta$ is a scalar
and $\basevec_i$ is the $i^{th}$ standard basis vector (in which only the
$i^{th}$ scalar is one).
In this case, $\sensemat\signal$ needs to be suitably updated (from
$\sensemat\signal_{old}$ to $\sensemat\signal_{new}$), which is to add the
$i^{th}$ column of $\sensemat$ (scaled by $\eta$) to the vector
$\sensemat\signal_{old}$.
When $\sensemat$ is sparse, the cost (time complexity) of this update is only
$O(\log(\belasize/\diffsize))$ (by \Cref{th:rip1}), whereas when $\sensemat$ is
dense, this cost is $O(\counternum) = O(\diffsize \log(\belasize/\diffsize))$,
which can be orders of magnitude larger.
In network measurement applications where $\sensemat\signal$ is maintained as a
(linear) {\it sketch} such as in~\cite{huang-nze}, and this sketch needs to be
updated upon every packet arrival, the $O(\counternum)$ update cost (when
$\sensemat$ is dense) can be too high for the packet stream on a high-speed
link (e.g., 100 Gbps).
Furthermore, even in applications where the encoding of $\signal$ is done by
multiplying $\sensemat$ to it in one shot, using a sparse $\sensemat$ can
reduce this encoding cost by a factor of $\diffsize$, as we will elaborate
in~\autoref{ssec:hash-encoding}.
Second, the computation cost of sparse recovery when $\sensemat$ is dense can
be at least $O(\diffsize / \log\belasize)$ times larger than when $\sensemat$
is sparse, according to\appendixref{sec:fast-omp}.

In fact, our \problemname{} solution is even pickier on $\sensemat$: We want
$\sensemat$ to be not only sparse, but also binary.
In this way, most scalars in $\sensemat\signal$ are small integers that take
few bits to store (and transmit), which achieves our goal of aggressively
compressing $\sensemat\signal$.
However, this ideal $\sensemat$ appears to contradict a recent negative result
established experimentally in~\cite{huang-nze}, which shows that such
$\sensemat$ is so ``far away'' from the RIP that correct sparse recovery is
infeasible using existing algorithms (such as OMP~\cite{tropp-greed-is-good}).
Since the work~\cite{huang-nze} also needs the matrix to be sparse (to maintain
a sketch), it proposes a compromise: a sparse $\sensemat$ with each nonzero
entry having a Gaussian distribution.
It is shown in~\cite{huang-nze} that this $\sensemat$ is ``slightly closer'' to
the RIP in the sense that it empirically leads to successful (nearly correct)
decoding but still has no theoretical guarantees.
However, this design does not work for our purpose, since each linear
measurement in the message $\sensemat\alisvec$ would be a real (floating point)
number, which takes several times more space to store (and transmit) than a
small integer that a binary sparse would afford us, and as a result cancels out
much of the hard-earned compression (ratio) that can be achieved by our
solution.

We solve this problem ($\sensemat$ being far away from the RIP) by discovering
and proving that the binary sparse $\sensemat$ used in \solutionname{}
satisfies a different RIP condition called RIP-1, a concept introduced
in~\cite{berinde-unified} that means approximate distance preservation in $L_1$
(Manhattan) distance.
Thanks to this RIP-1, sparse recovery can be done efficiently using a variant
of MP designed for the $L_1$ space: SSMP (sparse sequential matching
pursuit)~\cite{berinde-ssmp}.
We also discover that, since $\signal$ in our CS equation $\cscode =
  \sensemat\signal$ must be a binary vector, the vanilla MP
algorithm~\cite{mallat-matching-pursuit} designed for the $L_2$ space, which
runs faster than SSMP, can also correctly recover $\signal$ empirically.
Furthermore, we are the first to implement the data structures powering SSMP~\cite{berinde-ssmp}, to use them for speeding up the vanilla MP (with
nontrivial adaptions such as \Cref{def:us-update}), and to publish our source code and datasets at~\cite{common-sense-code}.
To the best of our knowledge, these data structures do not have a publicly available software
implementation as are typical for theory works like SSMP.


In summary, we make the following contributions in this paper.
\begin{enumerate}
  \item We debunk a longstanding perception that \problemname{} is as expensive
        as \reconname{} by showing a large gap between the information-theoretic
        lower bounds on the communication costs of these two problems.
  \item We propose \solutionname{}, the first exact \problemname{} protocol
        (that beats the lower bound of \reconname{}) without requiring $\alisset\subseteq \belaset$.
        Our protocol is also the first application of compressed sensing to this topic.
  \item We identify a CS matrix that is both friendly to applications and compliant to the CS theory
        (with guarantees on correctness and decoding time complexity).
        We show that our protocol achieves lower communication costs than the state of
        the arts using extensive experiments in~\autoref{sec:evaluation}.
\end{enumerate}

The rest of this paper is organized as follows.
In~\autoref{sec:motivation}, we describe five applications of \problemname{}.
Then, we describe our \solutionname{} protocol for unidirectional
\problemname{} (when $\alisset\subseteq \belaset$) and bidirectional
\problemname{} (without the above assumption) in~\autoref{sec:special-model}
and~\autoref{sec:model}, respectively.
We show how \solutionname{} can be adapted to a data streaming scenario
in~\autoref{sec:streaming}, and derive an information-theoretic lower bound on
the communication cost of \problemname{} in~\autoref{sec:with-set-recon}.
Finally, we compare the communication cost of \solutionname{} with
state-of-the-art protocols in~\autoref{sec:evaluation} and describe related
works in~\autoref{sec:related}.

\section{Applications of \solutionname{}}\label{sec:motivation}
In this section, we describe five potential applications of \problemname{}
that arise from various areas in computer science.
We first show an application of \emph{unidirectional} \problemname{} in
blockchain in~\autoref{sssec:blockchain}, followed by two data sketching
(streaming) applications in~\autoref{sssec:packet-loss}
and~\autoref{sssec:straggler}.
Then, we show two applications of \emph{bidirectional} \problemname{}
in~\autoref{sssec:antijoin} (relational database)
and~\autoref{sssec:delta-sync} (cloud file storage).
Recall that in unidirectional \problemname{} ($\alisset \subseteq \belaset$),
Bob computes $\belamalis$ (to obtain $\abintersec$).
This is also the case in bidirectional \problemname{} (when
$\alisset\not\subseteq \belaset$ and $\belaset\not\subseteq\alisset$), wherein
Alice, in addition, needs to compute $\alismbela$ (to obtain $\abintersec$).

All these applications can be solved by \solutionname{} using much less
communication cost or sketch size than by a \reconname{} protocol (say
invertible Bloom lookup table, or IBLT~\cite{Eppstein_WhatsDifference_2011}), since as
we have shown in our first contribution above, \problemname{} is fundamentally
much cheaper than \reconname{}.

\subsection{Propagation of New Blocks (Blockchain)}\label{sssec:blockchain}
Blockchains are distributed ledgers that often run on peer-to-peer computer
networks~\cite{Ozisik2019}.
In a typical blockchain system, new (unvalidated) transactions are often
relayed among peers upon generation, and they are temporarily stored in a
\emph{mempool}.
A special group of peers, called \emph{miners}, validate transactions and add
them into blocks that are certified by cryptographic
hashing~\cite{Merkle_merkletree_1987}.
A newly generated block needs to be propagated to the entire network and agreed
on by all peers under a \emph{consensus mechanism} before the transactions on
it are finalized.
Hence, efficient propagation of new blocks is crucial to achieving a high
throughput (transactions per second).

The Graphene paper~\cite{Ozisik2019} has shown that communication-efficient
propagation of new blocks can be reduced to unidirectional \problemname{} as
follows.
Suppose a peer named Alice just received a new block and wants to send it to
another peer named Bob.
Let the set of all transactions in that block be Alice's set $\alisset$, and
the set of all unvalidated transactions in Bob's mempool be his set $\belaset$.
In most cases (thanks to the aggressive relay of new transactions in most
blockchain implementations~\cite{Ozisik2019}), $\belaset$ is a superset of
$\alisset$.
As a result, Bob will find all transactions on the new block if he computes
$\abintersec$ ($=\alisset$) using a unidirectional \problemname{} protocol.
According to~\cite{Ozisik2019}, this approach uses much
less communication cost than a full-fledged \reconname{} protocol.

\subsection{Packet Loss Detection (Network
  Measurement)}\label{sssec:packet-loss}
Detecting packet losses in data center networks is crucial to diagnosing the
root causes of failures and mitigating their impacts.
LossRadar~\cite{li-lossradar} is a well-known algorithm to this problem.
In LossRadar, packet losses are detected by computing the difference between
the packets that traverse an upstream meter (switch) and a downstream meter,
with each packet identified by its signature consisting of a 5-tuple flow ID
and a packet ID which is consecutive within its flow.
Each switch along the path, in the data plane, digests all traversing packets
into an invertible Bloom lookup table (IBLT)~\cite{Eppstein_WhatsDifference_2011}.
Then, packet losses are identified in the control plane by comparing the
differences of the two digests (IBLTs).
A key performance metric of this problem is the size of the digest, since
high-speed data plane memory is a scarce commodity on programmable network
switches.
However, the IBLTs are not memory efficient, as we will show
in~\autoref{ssec:related-recon}.

We note that the streaming version of \solutionname{}, to be elaborated
in~\autoref{sec:streaming}, generates much leaner digests than IBLTs (to recover
the same number of lost packets), since the detection of packet losses between
two switches can be solved by a \problemname{} protocol as follows.
Treating the downstream meter as Alice and the upstream meter as Bob, with the
packets traversing them denoted by $\alisset$ and $\belaset$ respectively
($\alisset\subseteq \belaset$), the packets lost in transmission are those in
$\belaset \setminus \alisset$, which Bob computes in a \problemname{} protocol
as mentioned in~\autoref{ssec:second-contrib}.


Attentive readers who have read~\autoref{sec:streaming} may realize that
\solutionname{}'s decoder in the data streaming scenario requires a superset
$\belaset' \supseteq \belaset$ of all packets along this path.
Fortunately, $\belaset'$ can be readily collected by only incurring a small
overhead on the first switch on the path for the following reasons.
The set of all 5-tuple flow IDs can be recorded using existing methods such as
FlowRadar~\cite{li-flowradar}, and it is not hard to conservatively estimate
the range of packet IDs of each flow given that packet IDs are consecutive.


\subsection{Straggler Identification (Data Stream
  Processing)}\label{sssec:straggler}
Identifying the \emph{stragglers} in a data stream is an important research
problem motivated by the need to track outstanding resource allocations with a
minimum working set size~\cite{Eppstein_StragglerIdentificationRound_2011}.
To describe this problem, imagine that a library manages a catalog of millions
of books using a single computer that is \emph{very fast but has limited
  memory}.
Many people visit this library each day, who generate a data stream consisting
of millions of borrowing and returning records.
It is stated in the rules that all borrowed books must be returned by the end
of the day, and most (but regrettably not all) visitors abide by the rules.
The staffs want to know, at the end of a day, the set of books that are not
returned yet (i.e., the stragglers).
This is a nontrivial data streaming problem, since their
computer does not have enough memory either to load the entire catalog or to
track all outstanding books (borrowed but not returned) during the day.

The state-of-the-art solution is to keep an IBLT in
memory~\cite{Eppstein_StragglerIdentificationRound_2011}.
Similarly as above, this IBLT can also be replaced by a much smaller digest from
the streaming version of \solutionname{} as follows.
Denote by $\alisset$ the set of outstanding books.
In~\autoref{sec:streaming}, we will show that \solutionname{} can track any
addition to $\alisset$ (new borrows) or deletion from it (new returns)
efficiently using a small digest.
Given the catalog of books (the aforementioned $\belaset'$), the final state of
$\alisset$ can be losslessly recovered from the digest.


\subsection{Antijoin Computation (Relational Database)}\label{sssec:antijoin}
\solutionname{} can also lead to communication-efficient computation
of antijoin, an operator in relational database defined as
follows~\cite{galindo-outer-join}.
\begin{definition}
  Given two relations (tables) $R_1$ and $R_2$ with a common attribute (column)
  $S$, the antijoin $R_1 \triangleright R_2$ on $S$ consists of tuples (rows)
  $t\in R_1$ such that $t.S$ does \emph{not} appear in column $S$ of $R_2$.
\end{definition}

In this scenario, two relations $R_1$ and $R_2$ are stored at two hosts named
Alice and Bob respectively.
This is a common design in distributed databases called \emph{naive table
  partitioning in a shared-nothing architecture}~\cite{silberschatz2011database}.
The task is for Alice to compute $R_1 \triangleright R_2$ using as small a
communication cost as possible.

Without loss of generality, we assume that column $S$ of $R_1$ (denoted by
$\prod_{S}(R_1)$ in relational algebra) does not contain duplicate values
(i.e., $S$ is a \emph{candidate key} of $R_1$) and the same for $R_2$.
In this way, $\prod_{S}(R_1)$ can be regarded as the set $\alisset$,
$\prod_{S}(R_2)$ similarly as the set $\belaset$, and $R_1 \triangleright R_2$
consists of tuples that correspond to the elements in $\alismbela$, which is
computed by Alice in bidirectional \problemname{} as mentioned above.

\subsection{Matching Changes for Delta Synchronization (Cloud File Storage)}
\label{sssec:delta-sync}
File synchronization is an essential functionality of cloud storage services
such as Dropbox~\cite{dropbox}.
To save communication bandwidth, most cloud storage services adopt a
\emph{delta synchronization} mechanism, in which only the modifications (i.e.,
\emph{delta}) are transmitted between the server and the clients.
A typical delta synchronization protocol, such as rsync~\cite{tridgell-rsync},
works as follows.
During preprocessing, all stored files are cut into multiple \emph{chunks} so
that most user edits only involve one or two chunks\footnote{Although
  conventional fixed-size chunks can only handle replacement edits, recent
  \emph{content-defined chunking}~\cite{xia-netsync} techniques also handle
  insertions and deletions (see Figure~3 of~\cite{xia-netsync}).
}.
Then, in the \emph{matching} stage, the server and the clients identify all
deltas (modified chunks) by transmitting and comparing the checksums of all
chunks.
This consumes considerable network bandwidth, because the same checksums may be
transmitted frequently when a file is under active edits~\cite{xia-netsync}.
Finally, the detected deltas are resolved, and every participant reconstructs
the latest version of these files.

The problem of identifying deltas is actually equivalent to bidirectional
\problemname{} as follows.
Without loss of generality, consider a client named Alice who is making edits
and a server named Bob who is synchronizing these edits.
We denote the set of Alice's and Bob's chunks by $\alisset$ and $\belaset$,
respectively.
In the matching stage, Alice's goal is to compute $\alisset\setminus \belaset$,
her unique (recently modified) chunks to be sent to Bob, and Bob's goal is to
compute $\belaset \setminus \alisset$, his unique (obsolete) chunks to be
patched.
Both goals are achieved when solving \problemname{}.

In fact, many \reconname{} protocols also take file synchronization as a key
motivation~\cite{gong-pbs,yang-rateless-iblt}, but their contribution to this
problem is different from \problemname{} as follows.
The functionality of a \problemname{} protocol, as we just mentioned, is to
allow Alice to ``push'' her edits ($\alisset\setminus \belaset$) to Bob.
In comparison, a \reconname{} protocol also allows Bob to list these chunks
(that he misses from Alice), so that he can actively ``fetch'' such a chunk if
he needs it urgently.
However, this fetching functionality comes with a much larger communication
cost.

\section{Unidirectional Protocol: When $\alisset{} \subseteq \belaset{}$}\label{sec:special-model}

In this section, we describe our \solutionname{} protocol (the second
contribution of this paper) for unidirectional \problemname{} ($\alisset{}
\subseteq \belaset{}$).
We focus on unidirectional \problemname{} first, because its solution
(summarized in~\autoref{ssec:special-overview}) is more streamlined: The set
intersection can be computed in only one round of communication.
Nevertheless, unidirectional \solutionname{} contains all the essential
ingredients that make it a communication-efficient \problemname{} protocol,
such as compressed sensing (CS,~\autoref{ssec:compressed-sensing}), sparse
sensing matrices (\autoref{ssec:hash-encoding}) and a matching pursuit (MP)
decoder (\autoref{ssec:omp}).
For this reason, it serves as a nice starting point before we delve into a more
complex protocol for bidirectional \problemname{} in~\autoref{sec:model}.

\subsection{Protocol Overview}\label{ssec:special-overview}

We adopt the following vector representation of sets.
Any set $\aset$ in the universe $\universe \triangleq \{1,2,\ldots, 2^\bitwidth
  \}$ is represented by a $2^\bitwidth$-dimensional binary vector
$\indicator_{\aset}$, whose $i^{th}$ ($i=1,2\ldots, 2^\bitwidth$) scalar is $1$
if and only if $i\in\aset$.
Under this representation, Bob can achieve his objective in undirectional
\problemname{} (to compute $\belamalis$ and $\abintersec$) by computing
$\belavec - \alisvec$, since when $\alisset \subseteq \belaset$, it always
holds that $\indicator_\belamalis = \belavec - \alisvec$ and
$\indicator_\abintersec = \indicator_\alisset = \belavec -
  \indicator_\belamalis$.
Note that by our assumption ``around~\autoref{eq:assumption},'' $\belavec -
  \alisvec$ is a \emph{$\diffsize$-sparse} vector with only $\diffsize$ ones
among its $2^\bitwidth$ coordinates, wherein $\diffsize$ is the SDC
($|\alisset\triangle \belaset|$).
Alice does not need to compute anything in unidirectional \problemname{}, since
$\alisset$ is always equal to $\abintersec$.

The unidirectional \solutionname{} protocol is summarized in
\Cref{fig:uni-work}.
It applies the well-established \emph{compressed sensing} (CS)
technique~\cite{candes-rip} to the sparse vector $\indicator_\belamalis$, as
follows.
Let $\sensemat$ be an $\counternum\times2^{\bitwidth}$ \emph{CS matrix} as in
\Cref{def:sense-mat} (with $\numcols=2^\bitwidth$) and is shared between Alice
and Bob.
In step \textbf{1}, Alice compresses $\alisset$ into an
$\counternum$-dimensional vector (sketch) $\sensemat\alisvec$ and sends it to
Bob in the first and only round of communication in unidirectional
\solutionname{}.
Upon receiving this sketch (step \textbf{2}), Bob computes the
\emph{measurement} (of the \emph{sparse signal} $\indicator_\belamalis$)
$\cscode = \sensemat\indicator_\belamalis = \sensemat(\belavec - \alisvec) =
  \sensemat\belavec - \sensemat\alisvec$ from $\sensemat$, $\belaset$, and the
received $\sensemat\alisvec$.
Finally, he reconstructs $\indicator_\belamalis$ losslessly from $\cscode$
using an MP decoder in~\autoref{ssec:omp}, thereby achieving his objective in
unidirectional \problemname{} as just explained.

The communication cost and time complexities of this protocol are summarized as
follows.
\begin{theorem}\label{th:uni-main}
  \solutionname{} solves unidirectional
  \problemname{} using one round and $O(\diffsize\log(\belasize/\diffsize))$
  bits of communication (see the second remark in the last paragraph of~\autoref{ssec:hash-encoding}),
  $O(\belasize \log(\belasize/\diffsize))$ encoding time complexity for Alice
  (see~\autoref{ssec:hash-encoding}), and $O(\belasize(\log\belasize)
    \log(\belasize/\diffsize))$ decoding time complexity for Bob
  (by \appendixref{th:decode-time}).
\end{theorem}

Before elaborating on our designs of \solutionname{}, we show, using the
following example, that it achieves its \emph{primary} goal of minimizing the
communication cost and beating the lower bound of \reconname{}.

\begin{example}\label{exam:uni}
  In the first group of experiments in~\autoref{sssec:unidirecional-eval}, we let
  $\alisset\subseteq \belaset$ with $\alissize = 1,000,000$, $\belasize =
    1,010,000$ ($\diffsize = 10,000$), and $|\universe| = 2^{64}$.
  In this scenario, the information-theoretic lower bound on the communication
  cost of \reconname{}~\cite{Minsky2003} is roughly
  $\diffsize\log_2(e|\universe|/\diffsize)$ bits (or \SI{65.2}{KB}), whereas that
  of \problemname{} (by~\autoref{eq:csi-lb}) is only roughly
  $\diffsize\log_2(e\belasize/\diffsize)$ bits (or \SI{10.1}{KB}).
  The average communication cost of \solutionname{} measured in our experiments
  is only \SI{22.9}{KB} in this scenario, which is less than \reconname{}'s lower
  bound by a factor of 2.8.
  Moreover, suppose that the universe size $|\universe|$ becomes $2^{256}$ (as in
  the Ethereum dataset we mentioned in~\autoref{ssec:second-contrib}), it can be
  inferred that this outperformance factor (of \solutionname{}'s cost over
  \reconname{}'s lower bound) will grow to $13.3$ (since the former stays almost
  unchanged, whereas the latter becomes \SI{305.2}{KB}).
\end{example}



\subsection{Background
  on Compressed Sensing and a Strawman Protocol}\label{ssec:compressed-sensing}
In this subsection, we introduce some basic concepts of compressed sensing
(CS).
Based on these concepts, we can instantiate the overview
in~\autoref{ssec:special-overview} into a strawman protocol that
\emph{correctly solves} unidirectional \problemname{} using $O(\diffsize
  \log(\belasize/\diffsize))$ communication cost (but has high encoding and
decoding time complexities).

As mentioned in~\autoref{ssec:second-contrib}, CS offers a way to (almost)
losslessly reconstruct (recover) a high-dimensional sparse signal $\signal$
from a low-dimensional measurement of it, as follows.
Specifically, suppose the \emph{signal} $\signal$ is an $\numcols$-dimensional
vector with only $\diffsize$ ($\diffsize \ll \numcols$) nonzero coordinates
(i.e., $\signal$ is \emph{$\diffsize$-sparse}).
The \emph{measurement} $\cscode$ is an $\counternum$-dimensional vector
obtained by linearly transforming $\signal$ with an $\counternum \times
  \numcols$ \emph{CS matrix} $\sensemat$ and adding an $\counternum$-dimensional
\emph{noise} (vector) $\noise$ to it, as given by the following equation
\begin{equation}\label{eq:cs-encode} \cscode = \sensemat\signal + \noise.
\end{equation}

The following \emph{restricted isometry property} (RIP) is crucial to the
reconstruction quality of $\signal$.

\begin{definition}\cite{candes-rip,zhang-optimal-rip}\label{def:rip}
  A CS matrix $\sensemat$ has the RIP (for $\diffsize$-sparse signals) if for any
  $2\diffsize$-sparse vector $\vector$
  $$
    (1-\sqrt{2}/2)\cdot\|\vector\|_2^2 \le \|\sensemat\vector\|_2^2 \le
    (1+\sqrt{2}/2)\cdot\|\vector\|_2^2.
  $$
\end{definition}

\noindent According to~\cite{zhang-optimal-rip}, as long as $\sensemat$ has the RIP, the
following constrained $L_1$-optimization \footnote{In this paper, we denote the
  $L_p$ norm ($p=1, 2$) of a vector $\vector=(v_1,v_2,\ldots, v_m)$ by
  $\|\vector\|_p = \left(\sum_{i=1}^m|v_i|^p\right)^{1/p}$.
}
\begin{equation}\label{eq:cs-decode} \signal{}^*
  = \arg\min_{\signal}\|\signal\|_1, \; \mathrm{s.t.}
  \|\cscode -\sensemat\signal\|_2 < \epsilon
\end{equation}
($\epsilon$ is an upper bound on $\|\noise{}\|_2$, with an abuse of notation)
always computes a close approximation
$\signal^*$ to the original $\diffsize$-sparse signal $\signal$ such that
$\|\signal^* -
  \signal\|_2=O(\epsilon)$.
Furthermore, in the absence of noises ($\noise=\vec{0}$), any
$\diffsize$-sparse reconstruction $\signal^*$ (not necessarily from
$L_1$-minimization, as we will show in~\autoref{ssec:omp}) is always equal to
$\signal$ (i.e., a \emph{lossless reconstruction}) as long as $\cscode =
  \sensemat\signal^*$~\cite{candes-lp}, since $\signal$ is the unique
$\diffsize$-sparse solution to the above equation.


Although checking whether a given matrix has the RIP is
NP-hard~\cite{tillmann-rip}, many random constructions of RIP matrices have
been proposed.
They all, unsurprisingly, require the CS matrix to have a sufficient number of
rows so that the measurement $\cscode$ has a sufficiently high dimension to
losslessly represent the original signal.
For example, it has been shown that $\counternum \times \numcols$ random
Gaussian matrices are highly likely to have the RIP if $\counternum>
  \alpha\diffsize \log_2(\numcols/\diffsize)$ for some constant
$\alpha$~\cite{baraniuk-rip}.

With the backgrounds above, to instantiate \Cref{fig:uni-work} into a complete
(strawman) protocol, it suffices to specify that $\sensemat$ is an $\counternum
  \times 2^\bitwidth$ random Gaussian matrix (fixed upon generation) for some
$\counternum > \alpha \diffsize \log_2(\belasize/\diffsize)$ and that Bob's
reconstruction of $\indicator_\belamalis$ is by the $L_1$-minimization
in~\autoref{eq:cs-decode} while restricting all coordinates $i\notin \belaset$
of $\signal^*$ to zeros (since $(\belamalis)\subseteq \belaset$).

The key to the \emph{correctness} of this protocol is the fact that
$\sensemat^\belaset$, the $\counternum \times \belasize$ submatrix of
$\sensemat$ whose columns are restricted to the elements in $\belaset$, has the
RIP with high probability.
Since Bob's measurement $\cscode= \sensemat \indicator_\belamalis$ is
noiseless, the $L_1$-minimization (restricted to coordinates in $\belaset$) can
always losslessly reconstruct $\indicator_\belamalis$ given that
$\sensemat^\belaset$ has the RIP~\cite{zhang-optimal-rip}.
As a result, Bob can compute the correct set intersection with high probability
using a communication cost of sending $\counternum=O(\diffsize
  \log(\belasize/\diffsize))$ real numbers (in $\sensemat\alisvec$).


Although the strawman protocol is correct and has a small communication cost,
its time complexity is too high to be practical.
To encode $\alisset$ into $\sensemat\alisvec$, Alice needs to perform a (dense)
matrix-vector multiplication, which takes $O(\counternum \belasize)
  =O(\diffsize\belasize\log(\belasize/\diffsize))$ time; to reconstruct
$\indicator_\belamalis$, Bob needs to solve an $L_1$-minimization, whose time
complexity is $O(\diffsize^{1.5}\belasize^2)$ using the interior point
method~\cite{boyd-convex-opt}.
Both time complexities are prohibitively high on large datasets (where
$\belasize$ is usually in millions and $\diffsize$ in tens of thousands).

Fortunately, during decades of research, many fast CS techniques have been
proposed, although some of them sacrifice the reconstruction quality for speed.
In the next two subsections, we will explore these techniques and make careful
design choices so that \solutionname{}, as a \problemname{} protocol, is
\emph{both fast and correct}.
We will focus on encoding techniques in~\autoref{ssec:hash-encoding} and
decoding techniques in~\autoref{ssec:omp}.


\subsection{Fast Encoding via Sparse Binary CS
  Matrix}\label{ssec:hash-encoding}
In this subsection, we describe a \emph{sparse binary} CS matrix $\sensemat$
that reduces the encoding time complexity of \solutionname{} by a factor of
$\diffsize$ (from $O(\diffsize\belasize\log(\belasize/\diffsize))$ to
$O(\belasize\log(\belasize/\diffsize))$).
This sparse $\sensemat$ is also essential to the low update time complexity of
the data streaming version of \solutionname{} in~\autoref{sec:streaming}.
However, it has been shown that sparse matrices cannot have the
RIP~\cite{chandar-negative}, which all conventional CS guarantees rely on.
Fortunately, in the literature, a series of research
papers~\cite{berinde-unified,berinde-ssmp,price-any-rip1} have developed
theories and algorithms for \emph{sparse} $\sensemat$'s (that are parallel to
conventional ones based on the RIP).
For example, the seminal work~\cite{berinde-unified} has shown that sparse
matrices can have the RIP-1 as defined in \Cref{def:rip1} (instead of the RIP),
which can be viewed as the counterpart of the RIP in $L_1$ metric.
Indeed, \Cref{def:rip1} is essentially the same as \Cref{def:rip}, except for
the metric ($L_1$ norms instead of $L_2$) and different constant values.




\begin{definition}\label{def:rip1}\cite{berinde-unified}
  A (sparse) CS matrix $\sensemat$ has RIP-1 (for $\diffsize$-sparse signals) if
  for any $2\diffsize$-sparse vector $\vector$: 
  $$
  2/3\cdot \|\vector\|_1 \le \|\sensemat \vector\|_1 \le 4/3\cdot\|\vector\|_1.
  $$
\end{definition}
According to~\cite{berinde-unified}, all the guarantees on reconstruction
errors under the RIP in \autoref{ssec:special-overview} (to be precise, all
$L_2$ norms therein need to be replaced by $L_1$ norms) still hold under RIP-1.
Most importantly, in the absence of noises ($\vec{\epsilon} = \vec{0}$), any
$\diffsize$-sparse reconstructed signal $\signal^*$ is always \emph{lossless}
($\signal^*=\signal$) as long as $\cscode = \sensemat\signal^*$ and $\sensemat$
has the RIP-1.

In \solutionname{}, the binary matrix $\sensemat$ can be viewed as the
\emph{adjacency matrix} of a sparse \emph{random $\fanout$-right-regular
  bipartite graph} with $\counternum = O(\diffsize\log(\belasize/\diffsize))$
left nodes and $2^\bitwidth$ right nodes by \Cref{def:sense-mat}.
\begin{definition}\label{def:sense-mat}
    The adjacency matrix of a \emph{random $\fanout$-right-regular bipartite graph}, 
    where each $\numcols$ right node has $\fanout$ random neighbors
    from the $\counternum$ left nodes, is the following $\counternum \times \numcols$
    binary (zero-one) matrix $\sensemat$. 
    For each $i=1,2,\ldots, \numcols$, the $i^{th}$ column
    of $\sensemat$ (denoted by $\sensecol_i$) consists of \emph{ones} at
    $\fanout$ random coordinates (all fixed upon generation) and \emph{zeros} at
    all remaining coordinates.
\end{definition}

Note that in large universes (say $|\universe|=2^{256}$), it is impractical to
explicitly generate and store all $2^\bitwidth$ columns of $\sensemat$.
As a result, \solutionname{} adopts the following implicit pseudo-random
construction of $\sensemat$.
Suppose $h(\cdot)$ is a hash function (with fixed seeds) that maps
$i=1,2,\ldots, 2^\bitwidth$ to uniform random integers between 1 and
$\binom{\counternum{}}{\fanout{}}$, and $g(\cdot)$ is a one-to-one mapping from
integers on the above range to $\counternum{}$-dimensional binary vectors
containing exactly $\fanout{}$ ones.
For each $i=1,2,\ldots, 2^\bitwidth$, $\sensecol_i$ is defined as $g(h(i))$.
This matrix $\sensemat$ is shared between Alice and Bob as long as they share
$h(\cdot)$ (and its seeds) and $g(\cdot)$.

Thanks to the sparsity of $\sensemat$, the \emph{encoding time complexity} for
Alice to compute $\sensemat \alisvec$ (or for Bob to compute $\sensemat
  \belavec$) is only $O(\fanout \belasize)$, or
$O(\belasize\log(\belasize/\diffsize))$ as claimed by \Cref{th:uni-main} since
$\fanout = O(\log(\belasize/\diffsize))$ by \Cref{th:rip1}.
The same time complexity also holds under the implicit construction, as long as
the time complexities of evaluating $h(\cdot)$ and $g(\cdot)$ are both
$O(\fanout)$ (which are simple to achieve).

To show that Bob's encoded measurement $\sensemat\indicator_\belamalis$
contains \emph{enough information} for $\indicator_\belamalis$ to be losslessly
reconstructed, it suffices to show that the submatrix $\sensemat^\belaset$ has
the RIP-1 with high probability.
Note that $\sensemat^\belaset$ can also be regarded as the adjacency matrix of
a random $\fanout$-right-regular bipartite graph that has $\counternum$ left
nodes and $\belasize$ right nodes.
As a result, our target argument follows from the following theorems.

\begin{theorem}~\cite{berinde-unified} \label{th:expander}
  A matrix $\sensemat$ has the RIP-1 in \Cref{def:rip1} if it is the adjacency
  matrix of an \emph{expander} bipartite graph in the following sense: For any
  set $\aset$ of at most $2\diffsize$ right nodes, there are at least $5/6\cdot
    m|\aset{}|$ distinct left nodes that are neighbors to right nodes in $\aset$.
\end{theorem}
\begin{theorem}[Theorem 16 in~\cite{berinde-thesis}]\label{th:rip1}
  There exist parameters $\counternum = O(\diffsize\log(\belasize/\diffsize))$
  and $\fanout=O(\log(\belasize/\diffsize))$ so that a random bipartite graph as
  in \Cref{def:sense-mat} ($\numcols=\belasize$) is an expander in
  \Cref{th:expander} with high probability.
\end{theorem}

We conclude this subsection with two remarks on Alice's sketch
$\sensemat\alisvec$.
First, under \Cref{def:sense-mat}, this sketch can be regarded as a counting
Bloom filter (CBF) of $\alisset$~\cite{fan-cbf}
(see~\autoref{ssec:membership}), because the sketches of these two schemes, as
random vectors, follow the same distribution.
However, this is more of a coincidence than a deliberate act from our
perspective, because CBFs have never been used for CS purposes to the best of
our knowledge (there was an approximate \problemname{} protocol based on
CBF~\cite{Guo_SetReconciliationvia_2013} though, which we will elaborate
in~\autoref{ssec:setx-related}).
Second, as a result of $\sensemat$ being binary, this sketch is an
integer-valued vector, so it can be losslessly compressed (without quantization
errors) to a small size before transmission.
In fact, $\sensemat \alisvec$ can be compressed into
$O(\diffsize\log(\belasize/\diffsize))$ bits (the communication cost claimed by
\Cref{th:uni-main}) using entropy coding (whose details are deferred
to~\appendixref{sec:compress}).
In comparison, real-valued sketches such as the one in the strawman protocol
have a much higher communication cost.

\subsection{Fast and Capable Matching Pursuit (MP) Decoder}\label{ssec:omp}

In this subsection, we describe a fast decoding algorithm for Bob to
reconstruct $\indicator_\belamalis$ in $O(\belasize(\log\belasize) \\
  \log(\belasize/\diffsize))$ time (we defer the proof of the time complexity
to\appendixref{sec:fast-omp}, since it mainly follows the idea
in~\cite{berinde-ssmp}).
As a result, the decoding of \solutionname{} is much faster than the
aforementioned $L_1$-minimization and that of all ECC- (error correction
coding) based \reconname{} protocols (see~\autoref{ssec:related-recon}) and is only
slightly slower than that of IBLT~\cite{Eppstein_WhatsDifference_2011} (whose
time complexity is $O(\belasize)$) by two logarithmic factors.

Also, we show the \emph{correctness} of \solutionname{} in the following two
senses.
First, our decoder is \emph{capable} of reducing the residue (the difference
between $\cscode=\sensemat\indicator_\belamalis$ and $\sensemat\signal$, where
$\signal$ is the reconstructed signal) to zero empirically with high
probability.
Second, when that ($\cscode=\sensemat\signal$) happens, we can prove, via the
RIP-1 of $\sensemat$, that the reconstructed signal computes the \emph{exact}
set intersection (i.e., $\signal = \indicator_\belamalis$).

\begin{algorithm} \caption {Matching pursuit (MP)
    algorithm~\cite{mallat-matching-pursuit}.}\label{algo:omp}
  \KwIn{Measurement $\cscode_0$ and CS matrix $\sensemat$.}
  \KwOut{Reconstructed signal $\signal$.}
  Signal $\signal\gets \vec{0}$, and residue $\cscode\gets \cscode_0$
  \label{line:1}\; \While{$\|\cscode\|_2\ge \epsilon$ ($\epsilon$ is an upper
    bound on the $L_2$-norm of noise)\label{line:2}}{ $i^*, \delta^*\gets
      \arg\min_{i\in\universe, \delta\in\mathbb{R}}\|\cscode -
      \delta\sensecol_i\|_2$.
    (Matching)\;
    $\sigcor_{i^*} \gets \sigcor_{i^*} + \delta^*$ and $\cscode{} \gets
      \cscode
      - \delta^*\sensecol_{i^*}$,
    where $\sigcor_{i^*}$ is the ${i^*}^{th}$ coordinate of $\signal{}$
    (Update)\label{line:update}\;
  }
  return $\signal$\;
\end{algorithm}

Our decoder is based on \emph{matching pursuit}
(MP)~\cite{mallat-matching-pursuit}, a classical algorithm dating back to 1990s
that works for RIP (in $L_2$) matrices.
It is a \emph{nontrivial design choice} to use MP in \solutionname{}, since on
the one hand, the vanilla MP decoder is known to be less capable than
$L_1$-minimization, especially for (sparse) RIP-1 matrices that we use here.
Furthermore, as shown in~\cite{huang-nze}, a successor to MP called OMP
(\emph{O} stands for orthogonal) produces large reconstruction errors even when
the sketch dimension $\counternum$ is much higher than the RIP-1 bound (the
minimum $\counternum$ for $\sensemat$ to have the RIP-1).
On the other hand, we can certainly use one of the several MP-like decoders
proposed specifically for RIP-1 matrices~\cite{indyk-emp, berinde-smp,
  berinde-ssmp}.
However, they are more expensive computationally than vanilla MP decoders
(which we will explain in\appendixref{sec:example}).

Despite these negative results, we discover that the vanilla MP decoder is
\emph{surprisingly capable} of reconstructing {\it binary signals} (such as
$\indicator_\belamalis$), even when $\sensemat$ is RIP-1 (rather than RIP in
$L_2$ as normally needed by MP): According to our experiments
in~\autoref{sssec:unidirecional-eval}, it losslessly (correctly) reconstructs
$\indicator_\belamalis$ all the times (among 10,000 attempts) when
$\counternum$ is parameterized according to the RIP-1 theory
in~\autoref{ssec:hash-encoding}.
This discovery allows \solutionname{} to use the faster MP decoder (for RIP in
$L_2$) while ``provably enjoying'' the correctness guarantee afforded by the
RIP-1 theory, in solving the SetX problem.

Before elaborating our discovery, we first introduce the (vanilla) MP decoder.
As described in \autoref{algo:omp}, this algorithm iterates between two stages
(matching and update).
In each iteration, it selects a coordinate $i^*$ of the signal $\signal$ being
decoded so that pursuing this coordinate (by an optimal step $\delta^*$)
greedily leads to the minimum $L_2$ residue error $\|\cscode -
  \delta^*\sensecol_{i^*}\|_2$.
Correct pursuits (in which $\delta^*$ is equal to the ground truth signal on
coordinate $i^*$), are crucial to lossless reconstruction, especially in
earlier iterations, because pursuit errors are ``equivalent'' to noises
(see~\autoref{eq:alis-noise} for an example), and the MP decoder will stop
making progress if such noises accumulate to a high level.

In \solutionname{}, the signal to reconstruct is always binary, and only the
coordinates in $\belaset$ may be ones (The same invariant will also hold for
bidirectional \problemname{} in~\autoref{sec:model}).
To conform to this invariant, the update step in
\autoref{algo:omp} needs to be suitably modified as follows.
\begin{modification}\label{def:us-update}
  In \solutionname{}'s decoder, the signal $\signal$ is updated according to the
  following rules: 
  \begin{enumerate}[noitemsep, topsep=0pt]
    \item If $i^*\in\belaset$, $x_{i^*} = 1$ and
          $\delta^* < -1/2$, then $x_{i^*} \gets 0$ and $\cscode{} \gets \cscode{} +
            \sensecol_{i^*}$.
          \label{onetozero}
    \item	If $i^*\in\belaset$, $x_{i^*} = 0$ and $\delta^* > 1/2$, then
          $x_{i^*} \gets 1$
          and
          $\cscode{} \gets \cscode{} - \sensecol_{i^*}$.
    \item Otherwise, no update happens.\label{ignore}
  \end{enumerate}
\end{modification}


Now, we show the aforementioned correctness of \solutionname{}.
According to our experiments in~\autoref{sssec:unidirecional-eval}, this
decoder can reduce the residue $\cscode$ to $\vec{0}$ with high probability
empirically when $\sensemat$ is the aforementioned sparse binary matrix with
$\counternum = O(\diffsize\log(\belasize/\diffsize))$ rows.
Once $\cscode = \vec{0}$, Bob can \emph{correctly} recover $\belamalis$ as the
set of nonzero scalars in the reconstructed $\signal$ and then $\abintersec =
  \belaset \setminus (\belamalis)$, the \emph{correctness} of which is guaranteed
by the RIP-1 of $\sensemat^\belaset$ as mentioned
in~\autoref{ssec:hash-encoding}.
Finally, in the (rare in practice) case that the MP decoder fails to arrive at
a zero residue, Bob can fall back to one of the aforementioned MP decoders for
RIP-1, say sequential sparse matching pursuit (SSMP)~\cite{berinde-ssmp}.

According to~\cite{price-any-rip1}, SSMP, which is \emph{deterministic}
, guarantees to losslessly reconstruct $\indicator_\belamalis$ as long
as $\sensemat$ satisfies RIP-1 (although the specific definition of RIP-1
in~\cite{price-any-rip1} requires a larger $\counternum$ by a constant factor).

Finally, we remark that the significant boost of MP's reconstruction capability
on binary signals (we defer an intuitive explanation of it
to\appendixref{sec:example}) is an unexpected discovery.
Despite the high empirical reconstruction capability of our MP decoder
(compared with the negative results on non-binary signals in~\cite{huang-nze}),
to the best of our knowledge, only one work in the CS literature, named binary
matching pursuit (BMP)~\cite{wen-bmp}, focuses on binary signals.
BMP is based on a similar idea as ours.
However, it only allows zero-to-one updates but not the other way around (i.e.,
missing \autoref{onetozero} in \Cref{def:us-update}), so it cannot correct
erroneous updates made in previous iterations.
In comparison, our decoder can update $\vec{x}$ in both ways, which is crucial
to a lossless reconstruction especially for bidirectional
\problemname{} as will be shown in~\autoref{sec:model}.

\section{\solutionname{} on Data Streams}\label{sec:streaming}
In this section, we describe how \solutionname{} can be adapted to data
streaming applications such as those described in~\autoref{sssec:packet-loss}
and~\autoref{sssec:straggler}.
This adaptation addresses the following algorithmic aspects in which a typical
data streaming scenario differs from the formulation of \problemname{}
in~\autoref{sec:intro}.
\begin{enumerate}[topsep=1pt,itemsep=1pt]
  \item The sets $\alisset$ and $\belaset$ are not directly given; instead,
        their
        elements (or deletions of their elements as
        in~\autoref{sssec:straggler})
        arrive in a data stream.
        The time complexity of operating on each element needs to be low.
        \label{i1-stream}
  \item Instead of a small communication cost, the primary goal is low memory
        usage (small sketch size)
        when processing the data streams.\label{i2-stream}
  \item The sketch is usually decoded offline (after
        both data streams are processed), and in general, the decoder only
        knows a predetermined
        superset $\belaset'$ of $\belaset$ (since recording $\belaset$ also
        uses too much
        memory).\label{i3-stream}
\end{enumerate}

\noindent To address \autoref{i2-stream}, the streaming version of \solutionname{} stores
only the measurement $\cscode$ in memory, which is initialized to all zeros.
Upon the addition (or deletion) of an element $i$, $\sensecol_i$ is added to
(or subtracted from) $\cscode$ (which addresses \autoref{i1-stream}).
In this way, we can still get $\sensemat \indicator_\belamalis$ by subtracting
Alice's measurement from Bob's and then reconstruct $\indicator_\belamalis$
losslessly using the MP decoder (with $\belaset$ replaced by $\belaset'$
as a result of \autoref{i3-stream}).
For this reason, we need to substitute $\belasize$ with
$|\belaset'|$ when stating all the time and space complexity results in
\Cref{th:uni-main}.
That is, the space usage of storing the measurement is $O(\counternum)=
  O(\diffsize\log(|\belaset'|/\diffsize))$, the time complexity of updating each
element is $O(\fanout)= O(\log(|\belaset'|/\diffsize))$, and the time
complexity for decoding $\belamalis$ is $O(|\belaset'|(\log
  |\belaset'|)\log(|\belaset'|/\diffsize))$.

\section{Bidirectional Solution}\label{sec:model}

In this section, we extend \solutionname{} into a bidirectional \problemname{}
protocol that continues to work when neither set is a subset of the other (i.e.,
$\alisset \not\subseteq \belaset$ and $\belaset \not\subseteq \alisset$).
In this case, Alice needs to recover all her \emph{unique} elements
$\alismbela$ to arrive at $\abintersec = \alisset \setminus (\alismbela)$, and
similarly, Bob needs to recover $\belamalis$.
To this end, Alice and Bob perform \emph{ping-pong decoding}\footnote{Although
  the term \emph{ping-pong decoding} is borrowed from the Graphene
  paper~\cite{Ozisik2019}, our protocol is completely different.
} (\autoref{ssec:ping-pong}) as follows.
Alice first sends $\sensemat\alisvec$ to Bob like in the unidirectional case
above.
Bob then computes $\sensemat\belavec - \sensemat\alisvec =
  \sensemat\indicator_\belamalis - \sensemat\indicator_\alismbela$ (note the RHS
is not just $\sensemat\indicator_\belamalis$ as in the unidirectional case) and  
tries to reconstruct the \emph{signal component} ($\indicator_\belamalis$) therein using
the MP decoder.  Then, the residue (to be formally defined shortly) is sent back to Alice for further reconstruction.  
In this way, the residue ping-pongs between Alice and Bob until it becomes $\vec{0}$.
A zero residue guarantees that both Alice and Bob arrive at the correct set intersection
(to be proved in~\autoref{ssec:ping-pong}) provided that all so-called
\emph{common hallucinations} are avoided (to be described in~\autoref{ssec:fingerprint}).


Should Bob's reconstruction (from $\sensemat\indicator_\belamalis - \sensemat\indicator_\alismbela$) be completely correct (in precisely recovering $\indicator_\belamalis$), 
the residue would be precisely ($-\sensemat \indicator_\alismbela$).
Then, sending this residue back to Alice would let her recover $\indicator_\alismbela$ precisely (if $\sensemat$ satisfies RIP-1), 
since in this case Alice is faced with the same reconstruction problem as in a hypothetical ``reversed" (with $\belaset\subseteq \alisset$ and Bob 
initiating the protocol) unidirectional case.  In other words, in this (ideal) case, a single ping-pong exchange between Alice and Bob is enough.

In reality, however, the following more realistic scenario usually happens.  
Since ($-\sensemat \indicator_\alismbela$) is purely \emph{noise} to Bob as will be explained shortly, Bob usually cannot 
reconstruct the signal ($\indicator_\belamalis$) from the ``mixture" ($\indicator_\belamalis - \sensemat\indicator_\alismbela$) in a completely
correct manner in the following sense:  Bob can usually reconstruct ``most of" its signal ($\indicator_\belamalis$), while making some (false positive and negative) errors.  
As a result, the residue sent back to Alice would be ($-\sensemat \indicator_\alismbela$) plus (the ``effect of") these errors.  
Since Bob's signal ($\indicator_\belamalis$) is purely noise to Alice, Bob's (mostly correct) reconstruction significantly reduces this noise to just these errors, which in turn makes it much 
easier for Alice to reconstruct most of her signal ($-\sensemat \indicator_\alismbela$) in the next step.  
As such, the actions of Alice and Bob in this ping-pong decoding process are mutually beneficial:  Bob's reconstruction (of his 
signal) significantly reduces noise to Alice's reconstruction, and vice versa.
  
While this mutually beneficial interaction explains how the ping-pong decoding 
process makes (often good) progress, it alone cannot explain why the process can usually successfully finish (by reducing the final residue to $\vec{0}$) after only a few ping-pong rounds 
empirically.  The missing explanation is the \emph{noise-resilience} property of MP
decoders:  By design, an MP decoder can correct its \emph{own} (decoding) errors (in the past)
once the \emph{external} noise is removed or significantly reduced from the residue.
For example, as mentioned above, in the first round Bob would likely make some errors in reconstructing its signal ($\indicator_\belamalis$) in the presence of strong noise ($-\sensemat \indicator_\alismbela$); 
but the MP decoder allows Bob to correct, in the second round, most of his earlier errors, after Alice has significantly reduced this noise (which is Alice's signal).

The communication cost and time complexity of
our bidirectional protocol can be summarized as follows.
\begin{assumption}\label{obs:general}
  \solutionname{} empirically solves bidirectional \problemname{} in $R\le 10$
  rounds of communications, using $O(\diffsize
    \log(\belasize/\diffsize))$ bits of total communication cost
  (about twice the cost in the unidirectional case)
  and $O(\belasize (\log\belasize) (\log(\belasize/\diffsize)+R))$ total time
  complexity (see the paragraph after the proof of \appendixref{th:decode-time}).
\end{assumption}
\begin{example}\label{exam:bi}
  In the fourth group of experiments shown in~\autoref{sssec:bidirectional-eval}, we
  let $\alissize = \belasize = 1,010,000$, $\diffsize = 20,000$ ($|\alismbela| =
    |\belamalis| = 10,000$) and $|\universe|=2^{256}$.
  In this scenario, the information-theoretic lower bound on the communication
  cost of \reconname{}~\cite{Minsky2003} is roughly
  $\diffsize\log_2(2e2^\bitwidth/\diffsize)$ bits (or \SI{610.4}{KB}), whereas
  that of \problemname{} (by~\autoref{eq:csi-lb}) is only roughly
  $\diffsize\log_2(2e\belasize/\diffsize)$ bits (or \SI{20.3}{KB}).
  The average communication cost of \solutionname{} measured in our experiments
  is \SI{129.2}{KB} in this scenario, which is less than \reconname{}'s lower
  bound by a factor of 4.7.
\end{example}


\subsection{Ping-Pong Decoding}\label{ssec:ping-pong}
Recall that in~\autoref{eq:assumption}, we have named the host with a smaller
cardinality Alice and the other host Bob (hence $|\alismbela| < |\belamalis|$).
In bidirectional \solutionname{}, we always let Alice initiate the protocol,
because it leads to a smaller communication cost than the other way around, as
we will show shortly.


\Cref{fig:bi-work} shows the workflow of bidirectional \solutionname{}.
Its first two steps are the same as in the unidirectional case.
Specifically, in step \textbf{1}, Alice sends $\sensemat\alisvec$ to Bob.
Then in step \textbf{2}, Bob tries to decode $\cscode_{(1)}=\sensemat\belavec -
  \sensemat\alisvec= \sensemat \indicator_\belamalis - \sensemat
  \indicator_\alismbela$ (as signal coordinates in $\alisset \cap \belaset$
cancel out) using the MP decoder (described in~\autoref{ssec:omp}), where the
subscript ``(1)'' in $\cscode_{(1)}$ refers to $\cscode_{(1)}$ being derived
from the information contained in the \emph{first} message (and so do ``(2)'' and ``(3)'' later in the protocol).

In general, Bob is not able to fully decode
$\indicator_\belamalis$ from $\cscode_{(1)}$ for the following reason.
Unlike in the unidirectional case, $\cscode_{(1)}$ now has a component
($-\sensemat \indicator_\alismbela$) that is purely noise to 
Bob, since Bob can only decode coordinates in $\belaset$ as stated in
\Cref{def:us-update}, but we have $(\alismbela)\bigcap \belaset=\emptyset$.
This noise component interferes with Bob's decoding.
As a result, what Bob recovers from $\cscode_{(1)}$, which we denote as $\hbelamalis$, 
is likely different from the actual $\belamalis$.
$\hbelamalis$ can contain
both false positives 
that are precisely the aforementioned hallucinations, and false negatives.


In the second step, after subtracting the ``effects" of $\hbelamalis$ from $\cscode_{(1)}$ (by the
update rules in \Cref{def:us-update}), Bob obtains the following \emph{residue}
\begin{equation}\label{eq:alis-noise} \cscode_{(2)} \triangleq \cscode_{(1)} -
  \sensemat \indicator_\hbelamalis = \sensemat \indicator_\belamalis - \sensemat
  \indicator_\hbelamalis -\sensemat \indicator_\alismbela.
\end{equation}

From this point on, Alice and Bob start \emph{ping-pong decoding}, as shown in
\Cref{fig:bi-work}.
In step \textbf{3}, Bob sends the residue $\cscode_{(2)}$
to Alice.
Then in step \textbf{4}, Alice tries to decode $\indicator_\alismbela$ from
$\cscode_{(2)}$.
If this decoding is again imperfect (say due to the nonzero noise component
$\sensemat \indicator_\belamalis - \sensemat \indicator_\hbelamalis$
in~\autoref{eq:alis-noise}), Alice sends her residue $\cscode_{(3)}$ to Bob in
the \emph{third} message (step \textbf{5}).
In this way, the residue ping-pongs between Alice and Bob (shown as the
ellipsis in \Cref{fig:bi-work}) before it is reduced to $\vec{0}$, which
empirically always happens (in all our 10,000 experiments with $\counternum =
  O(\diffsize \log(\belasize/\diffsize))$) within ten rounds of communications.

Now we state a fact that explains why this ping-pong decoding always completes
in ten rounds empirically.
Here, with a slight abuse of notation, we let $\hbelamalis$ and $\halismbela$
denote what Bob and Alice believe to be (the contents of) $\belamalis$ and
$\alismbela$ after the last residue ($\cscode_{(t-1)}$) is decoded, respectively.
This fact holds because both Alice and Bob have subtracted the ``effects" of
$\halismbela$ and $\hbelamalis$ from $\sensemat\indicator_\alismbela$ and
$\sensemat\indicator_\belamalis$, respectively.

\begin{fact}\label{my-signal-your-noise}
  The following equality holds for all residues $\cscode_{(t)}$ in each message
  $t=3,4,\ldots$ \begin{equation}\label{eq:bob-noise} \cscode_{(t)} =
    \sensemat\indicator_\belamalis - \sensemat \indicator_\hbelamalis -
    (\sensemat\indicator_\alismbela - \sensemat\indicator_\halismbela).
  \end{equation}
\end{fact}


Note that $\sensemat\indicator_\belamalis - \sensemat \indicator_\hbelamalis$
encodes all the \emph{current} decoding errors of Bob (which is precisely the set
$(\belamalis)\triangle(\hbelamalis)$, wherein the false negatives are encoded with plus signs and
the false negatives (hallucinations) are encoded with minus signs).
They are purely noise to Alice, for the following reasons.
On the one hand, false negatives are a subset of $\belamalis$ (hence outside 
$\alisset$), which Alice is not able to decode as explained above.
On the other hand, Bob's hallucinations are a subset of $\hbelamalis$, which
Alice also does not decode as instructed by the SMF
(to be described in~\autoref{ssec:fingerprint}).
Similarly, we can show that Alice's decoding errors are purely noise to Bob.
Hence Bob, by decoding his signal from $\cscode_{(1)}, \cscode_{(3)},
  \cscode_{(5)}, \ldots$, reduces the noise (that Alice has to deal with) in
$\cscode_{(2)}, \cscode_{(4)}, \cscode_{(6)}, \ldots$ and vice versa.
It is this aforementioned mutually beneficial interaction 
that leads to the empirical success of ping-pong decoding.

Now we prove that, in the absence of common hallucinations (or equivalently
$(\hbelamalis) \cap (\halismbela) = \emptyset$, as we will show
in~\autoref{ssec:fingerprint}), when the ping-pong decoding completes, with
high probability Bob (resp., Alice) correctly computes $\belamalis$ (resp.,
$\alismbela$) and hence $\abintersec$ as shown in the preamble of this section.
As mentioned in~\autoref{ssec:hash-encoding}, as long as
$\sensemat^{\alisset \cup \belaset}$ (the submatrix of $\sensemat$ whose
columns are restricted to the elements in $\alisset \cup \belaset$) has the
RIP-1 (which holds with high probability if $\counternum = O(\diffsize
  \log(\belasize/\diffsize))$), a zero residue implies that
$\indicator_\hbelamalis - \indicator_\halismbela = \belavec-\alisvec =
  \indicator_\belamalis - \indicator_\alismbela$.
Moreover, since $(\belamalis)\cap(\alismbela) = \emptyset$, when $(\hbelamalis)
  \cap (\halismbela) = \emptyset$ (no common hallucinations), it must be that $\hbelamalis =\belamalis$ and
$\halismbela = \alismbela$.
That is, the reconstruction of both $\indicator_\alismbela$ and
$\indicator_\belamalis$ are lossless.

Finally, we explain why, under the aforementioned assumption $|\alismbela| <
  |\belamalis|$, it is better for Alice to initiate the protocol instead of Bob.
Consider the alternative scenario in which Bob initiates the protocol.
In this case, it is Alice who tries to decode $\cscode_{(1)}$, and $-
  \indicator_\alismbela$ is the signal from her perspective while $\sensemat
  \indicator_\belamalis$ is the noise.
Comparing these two scenarios (Alice being the initiator or Bob doing so), the
former case leads to a higher reconstruction quality (and a smaller communication
cost to reach lossless reconstruction eventually), because, under the assumption above,
its noise level
($\|\sensemat \indicator_\alismbela\|_1$) is statistically lower than the other
way around ($\|\sensemat \indicator_\belamalis\|_1$).

\subsection{Avoiding Common Hallucinations with Set Membership Filter
  (SMF)}\label{ssec:fingerprint}

In this subsection, we describe the aforementioned \emph{common hallucination}
problem, in which an element (say $i$) becomes ``hallucinatory'' to both
Alice and Bob.
When this happens, both Alice and Bob will regard $i$ as being unique to them,
even though by definition, $i$ belongs to the set intersection
$\abintersec$.
Hence, in order for Alice and Bob to compute the correct $\abintersec$, they
need to avoid all common hallucinations.
We describe one such solution using set membership filters (SMFs) in this
subsection.


Unfortunately, in the vanilla ping-pong decoding procedure shown
in~\autoref{ssec:ping-pong}, \emph{hallucinations are contagious}.
We describe this fact using an example, in which Bob first develops
a hallucination on an element $i\in\abintersec$ when trying to decode $\cscode_{(1)}$ (due to the noise 
component $-\sensemat \indicator_\alismbela$ therein).
As a result, he mistakenly subtracts $\sensemat\basevec_i$ ($\basevec_i$ is the
1-sparse binary vector whose only one is on the $i^{th}$ coordinate) from the
residue $\cscode_{(1)}$ according to \Cref{def:us-update}, so $\cscode_{(2)}$
has a ``signal'' component $-\sensemat\basevec_i$ in it (which is encompassed in the
$-\sensemat\indicator_{\hbelamalis}$ term in~\autoref{eq:alis-noise}).
When $\cscode_{(2)}$ is passed on to Alice, Alice considers $\basevec_i$ to be a 
part of her signal, because $i\in\alisset$ (which means she can decode $i$),
and $-\sensemat\basevec_i$ has the same minus sign as her actual signal
$-\sensemat\indicator_\alismbela$ in~\autoref{eq:alis-noise}.
For this reason, Alice will likely include $i$ in $\halismbela$, or
\emph{catching this hallucination from Bob}, which results in a common
hallucination.

The aftermath of such a common hallucination is that $i$ will be (mistakenly)
missing from the $\abintersec$ computed by both Alice and Bob, for the following
reason.
When Alice catches the hallucination on $i$, she also subtracts
$\sensemat\basevec_i$ from $\cscode_{(2)}$ by \Cref{def:us-update}.
However, since her signal has an opposite sign as Bob's, this substraction
\emph{cancels out} Bob's previous (mistaken) subtraction of $\sensemat\basevec_i$, so $i$ ``disappear" from the residue for good.
As a result, both Alice and Bob will ``happily'' accept $i$ as being unique to
them when the residue is reduced to zero.
Had Alice not developed the common hallucination on $i$, the signal $-\sensemat\basevec_i$ would remain
in the residue, which would allow Bob to eventually \emph{correct his hallucination} later in 
ping-pong decoding, thanks to the aforementioned error resilience of MP decoders.  

Before we state our fix to this problem, we highlight two aforementioned facts.
First, this problem does not affect the correct decoding of other ``innocent'' 
elements unique to Bob (or Alice).
Second, although having some elements missing from (the computed) $\abintersec$
is technically incorrect, it is innocuous in some applications.
For example, in file synchronization described in~\autoref{sssec:delta-sync},
it is acceptable to have a few hallucinations in the delta (since they will
eventually be detected and removed when deltas are resolved).

The fix to common hallucination is to enforce the invariant that $\hbelamalis$
and $\halismbela$ do not share any common element (or $(\hbelamalis)\cap(\halismbela)=\emptyset$ as mentioned
earlier), since such elements
are precisely common hallucinations.
To this end, Alice needs to know which elements have been added to
$\hbelamalis$ by Bob, and vice versa.
Here we only describe our solution for Alice, since that for Bob is
``symmetrically similar." In this solution, we let Bob encode $(\hbelamalis)$
into a \emph{set membership filter} (SMF, in~\autoref{ssec:membership}) and
sends this filter to Alice along with his residue.
Correspondingly, the MP decoder (\Cref{def:us-update}) at Alice's end will not update a signal
coordinate $i^*$ (selected in the matching stage) if $i^*$ tests positive in this filter.

So far, this solution is still incomplete, because all (implementations of) SMFs have \emph{false positives}.
If an element in $\alismbela$ is a false positive to the SMF, then it will not
add $\halismbela$ (and hence will be erroneously missing from the $\alismbela$ computed by Alice).
This ``false positive'' problem can be solved by a ``last inquiry'' step as
follows.
After a few rounds of communications, when Alice becomes confident that Bob has corrected almost all his
hallucinations (which is expected now with the use of SMF to prevent Alice from
joining Bob in these hallucinations), she can proactively verify with Bob whether
each positive element (judged by the SMF) actually belongs to $\alismbela$ or is Bob's
hallucination.
The communication cost of this ``last inquiry'' can be made very small, by
representing every positive element by a hash signature (that is large enough
to avoid ``birthday attacks''~\cite{flajo-birthday} in which any two elements in $\abunion$ are
hashed to the same value).

In summary, whenever a signal coordinate $i^*$ is selected by Alice's MP
decoder but tests positive in the SMF, she has the following two options.
\begin{enumerate}[itemsep=2pt,topsep=2pt]
    \item \textbf{Collision avoidance:} Initially, she ignores
        this coordinate (to give Bob a chance to correct his hallucinations) 
        and selects the next coordinate that minimizes the $L_2$ residue error.
      \item \textbf{Collision resolution:} When confident,
        she can first tentatively update these coordinates according to
        \Cref{def:us-update} and then verify these updates with Bob (through the ``last inquiry").
        All erroneous (hallucinatory) updates are then reverted.
\end{enumerate}
Our invariant ($\hbelamalis) \cap (\halismbela) = \emptyset$ always holds under
both options.
Therefore, as shown above, both Alice and Bob obtain the correct set
intersection once the residue is reduced to zero.

\section{Information-Theoretic Lower Bound of \problemname{}}\label{sec:with-set-recon}

In this section, we derive an information-theoretic lower bound on the
communication cost of the $\mathrm{\problemname}_\universe(\alisset, \belaset)$ problem
(computing the intersection between $\alisset$ and $\belaset$ in universe
$\universe$).
It is well-known in information theory that the communication cost is lowered
bounded by the \emph{reduction in entropy}.

In our derivation, we assume that both $|\alismbela|$ and $|\belamalis|$ are known
beforehand (so we do not charge the communication cost of computing these
values).
From Alice's perspective, the number of possible ways to partition $\alisset$
into $\abintersec$ and $\alismbela$ is $\binom{\alissize}{|\alismbela|}$, which
corresponds to $\log_2\binom{\alissize}{|\alismbela|}$ bits of entropy under a
uniform prior distribution.
Similarly, Bob's entropy is $\log_2\binom{\belasize}{|\belamalis|}$ bits a
priori.
After running a \problemname{} protocol, Alice and Bob know the exact set
intersection, so their entropies are both reduced to zero.
As a result, we have \begin{equation}\label{eq:csi-lb}\begin{split}
    \mathrm{Comm.
      \;Cost}[\mathrm{\problemname}_\universe(\alisset, \belaset)] &\ge
    \log_2\binom{\alissize}{|\alismbela|} + \log_2\binom{\belasize}{|\belamalis|}
    \\&\approx |\alismbela| \log_2(e\alissize/|\alismbela|) + |\belamalis| \log_2(e\belasize/|\belamalis|)
    \ge  \diffsize\log_2(e\alissize/\diffsize).
  \end{split}
\end{equation}

In~\autoref{eq:csi-lb}, the approximation ($\approx$) is by the Stirling's
formula, and the last inequality is by the facts $\alissize\le \belasize$, $|\alismbela|\le
  \diffsize$, and $|\belamalis|\le \diffsize$ ($\diffsize$ being the SDC).

Before moving on to evaluation, we remark that it is possible that the
information-theoretic lower bound in~\autoref{eq:csi-lb} cannot be achieved by
any practical protocol (say with a reasonable time complexity).
In fact, in theoretical computer science, the minimum communication cost
achievable by any (deterministic or randomized) protocol is called the
\emph{communication complexity}~\cite{comm-comp-book}.
So far, we do now know whether the communication complexity of \problemname{} (under a certain number of rounds of communications)
is equal to~\autoref{eq:csi-lb} or is strictly higher.
To the best of our knowledge, the most related result in the literature is on
\emph{set disjointness} (deciding whether $\alisset \cap \belaset$ is
empty)~\cite{hastad-disjointness, huang-sic-comm}, and its randomized
communication complexity (when $\alissize = \belasize$) has been shown to be
$\Omega(\belasize)$~\cite{hastad-disjointness}.
This value is much higher than the communication cost of \solutionname{} (a
randomized protocol), which is $O(\diffsize\log_2(\belasize/\diffsize))$ (by
\Cref{th:uni-main}) relying on a small $\diffsize$.
There is no contradiction between these two complexity values, however, because the former
result does not assume known $\diffsize$.
\section{Performance Evaluation}\label{sec:evaluation}
In this section, we evaluate the communication costs of our \solutionname{}
protocol on both synthetic data and a massive real life dataset (Ethereum
accounts) with more than 290 million elements.
The results show conclusively that \solutionname{} has much less communication
cost than existing protocols.
It beats Graphene~\cite{Ozisik2019} in unidirectional \problemname{} by a
factor of up to 7.4 (\autoref{sssec:unidirecional-eval}), and IBLT~\cite{Eppstein_WhatsDifference_2011}
in bidirectional \problemname{} by a factor of up to 14.8 (\autoref{sssec:bidirectional-eval}).
Moreover, a similar outperformance over IBLT (by about one order of magnitude)
is also observed on the Ethereum dataset (\autoref{ssec:ethereum}).

\subsection{Evaluation Setup}\label{ssec:setup}
\noindent\textbf{Datasets} Two types of datasets are used in our experiments.
In~\autoref{ssec:synthetic}, we use synthetic datasets in which $\alisset$ and
$\belaset$ consist of randomly generated elements.
For each group of parameters ($|\alisset|$, $|\belaset|$, and the SDC
$\diffsize$), we generate 10,000 pairs of $\alisset$ and $\belaset$ (instances)
using different pseudorandom number generator seeds.
In~\autoref{ssec:ethereum}, we use a massive real life dataset from the
Ethereum blockchain downloaded from the PublicNode website~\cite{publicnode}.
We will elaborate on this dataset in~\autoref{ssec:ethereum}.

\noindent\textbf{Algorithms and Parameters} For unidirectional \problemname{}, we compare \solutionname{} with
Graphene~\cite{Ozisik2019}, the state of the art for unidirectional
\problemname{} (see~\autoref{ssec:setx-related}).
For bidirectional \problemname{}, because there are no existing protocols,
\solutionname{} is compared with two \reconname{} protocols, namely IBLT
(invertible Bloom lookup table)~\cite{Eppstein_WhatsDifference_2011}, which is
popular and fast (see~\autoref{ssec:related-recon}), and ECC (error correction coding),
which is communication-efficient but slow (see~\autoref{ssec:related-recon}). 
The parameter settings of \solutionname{} and all competitors are as follows.

\begin{itemize}[itemsep=2pt,topsep=2pt]
    \item\noindent\textit{\solutionname{}}
    \solutionname{} has two parameters in its compressed sensing (CS) matrix
    $\sensemat$,
    namely $\counternum$ (the number of rows)
    and $\fanout$ (the number of ones in each column).
    We tune $\counternum$ in each group of experiments so that it is close to the
    minimum value under which a random instance is always losslessly reconstructed
    empirically.
    $\fanout$ is fixed at 7 for unidirectional \problemname{} and 5
    for bidirectional \problemname{}.
    
    \item\noindent\textit{Graphene}
    We use the Python library~\cite{graphene-github} shared by the authors
    of~\cite{Ozisik2019}.
    That library can automatically determine its parameters from three input
    values: $\alissize$, $\belasize$, and a probability bound $\beta$ on
    successfully decoding the IBLT (which we set to the default value of
    $239/240$).
    
    \item\noindent\textit{IBLT}
    We use the same C++ implementation of IBLT~\cite{iblt-github} as used in Graphene.
    We use a heuristic parameter setting of $\fanout=4$ hash functions, and a hedge
    factor of 1.36 (so that the number of cells $\counternum$ is roughly
    $1.36\diffsize$).
    Each IBLT cell has a fingerprint field that is 48-bit-long in the Ethereum
    experiment in~\autoref{ssec:ethereum} and 32-bit-long in all other experiments.
    Note that IBLT needs two rounds of communications to solve bidirectional
    \problemname{}.
    In the first round, Alice encodes $\alisset$ into an IBLT and sends it to Bob,
    and Bob, upon decoding the difference between two IBLTs, knows both $\alismbela$
    and $\belamalis$.
    In the second round, Bob encodes $\alismbela$ into a message of $|\alismbela|
      \log_2\alissize$ bits in length and sends it back to Alice.
    
    \item\noindent\textit{ECC}
    Since the decoding time complexity of ECC-based protocols is prohibitively
    high, instead of actually running an ECC-based protocol, we
    \emph{optimistically} (to our disadvantage) estimate its communication cost
    using the information-theoretic lower bound for \reconname{}
    in~\cite{Minsky2003}.
    Since these values are not from actual experiments, the corresponding line in
    \Cref{fig:bidirect} does not have markers.
\end{itemize}

\noindent
Specifically, we assume the SDC $\diffsize$ is known to all protocols, because
it can be handily estimated using min-wise hashing~\cite{Broder_minwise_1997},
Strata~\cite{Flajolet_Strata_1985}, tug-of-war sketch~\cite{gong-pbs}, or
GXBits~\cite{jia-gxbits}, by sending a few hundred bytes during a handshake
step.

\noindent\textbf{Evaluation metrics}
As mentioned at the end of~\autoref{sec:psea}, our primary design goal is to
minimize the communication cost.
Hence, for each group of experiments below, we report the total number of bytes
transmitted in all rounds of communications (i.e., the overall communication
cost) averaged over all instances in the group, under the requirement that the
\emph{correct (exact)} set intersection is computed in \emph{all} instances.

\noindent\textbf{System Setup}
We conduct all our experiments on a workstation running Ubuntu 22.04 with Intel
Core i9--10980XE \SI{3.0}{GHz} CPU, \SI{256}{GB} DRAM, and \SI{108}{GB} swap
space on a Micron 2300 NVMe 1024 GB SSD\@.
We compile C++ programs with g++-14.1 ``-O3'' optimization and
our Python version is 3.9.7.

\subsection{Results on Synthetic Data}\label{ssec:synthetic}
\noindent\textbf{Unidirectional \problemname{}}\label{sssec:unidirecional-eval}
In this experiment, we set the universe size $|\universe|$ to $2^{64}$,
corresponding to the 8-byte transaction IDs used in the Graphene
paper~\cite{Ozisik2019}.
We conduct eight groups of experiments: $|\alisset|$ is fixed at 1,000,000 in
all groups, and the number of Bob's unique elements $\diffsize = |\belamalis|$
ranges from 10,000 to 2,500,000.
For each group, 10,000 instances of $\alisset$ and $\belaset$ are randomly
generated such that $\alisset\subseteq \belaset$ in each instance.
We have ensured that exactly the same instances are generated across C++
(\solutionname{}) and Python (Graphene) programs.

\begin{figure}
  \begin{subfigure}[b]{0.47\textwidth} \centering
    \includegraphics[width=\textwidth]{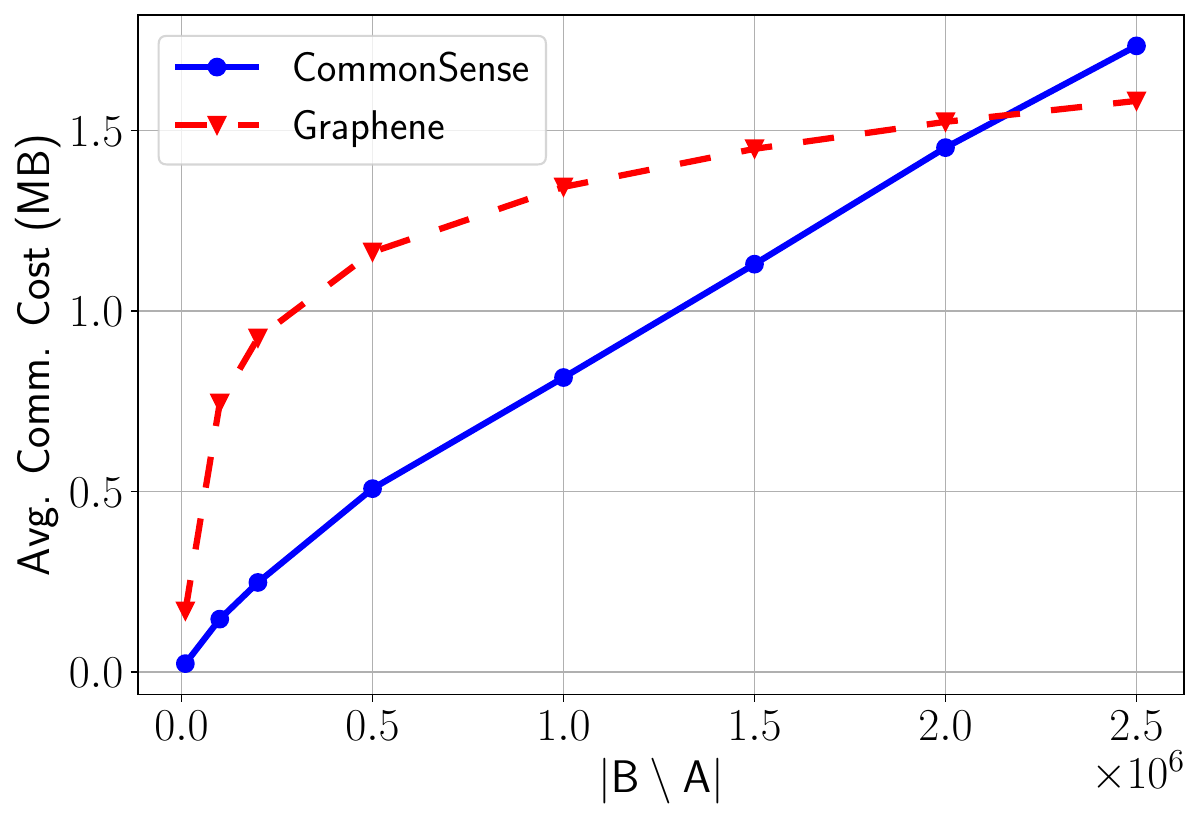}
    \caption{Unidirectional
      \problemname{} results as $|\belamalis|$ increases from 10,000 to
      2,500,000.
    }\label{fig:unidirect}
  \end{subfigure}
  \begin{subfigure}[b]{0.47\textwidth}
    \centering
    \includegraphics[width=\textwidth]{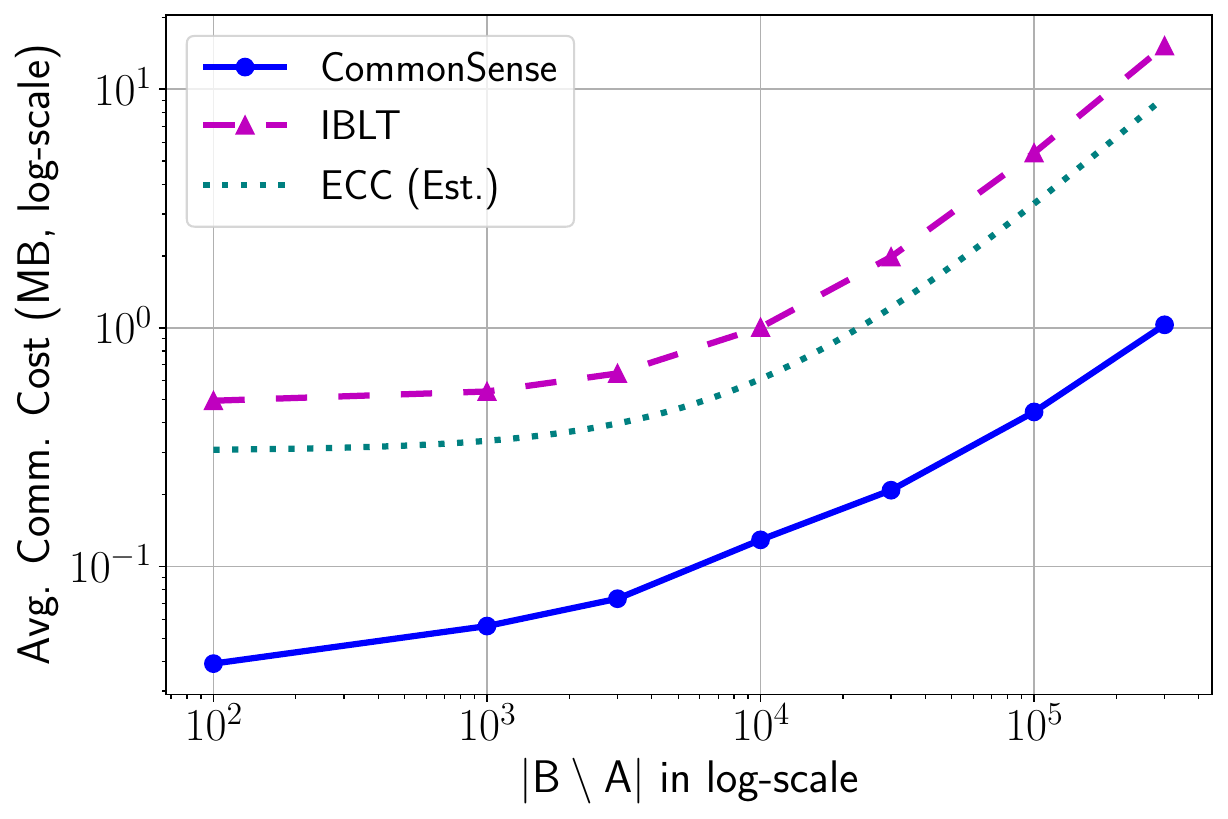}
    \caption{
      Bidirectional \problemname{} results as $|\belamalis|$
      increases from 100 to 300,000.
      $|\alismbela|$ is fixed at 10,000 in all groups.
      Both axes are plotted in log-scale.
    }\label{fig:bidirect}
  \end{subfigure}
  \caption{Comparison of average communication costs among 10,000 randomly
    generated
    pairs of $\alisset$ and $\belaset$ in each group.
    We fix $|\abintersec|$ at 1,000,000 in all groups.
    The communication costs of error correction codes (ECC) are optimistically (to
    our disadvantage) estimated from the lower bounds in~\cite{Minsky2003}.
    \SI{1}{MB} is $10^6$ bytes.}\label{fig:main}
\end{figure}

\Cref{fig:unidirect} shows that \solutionname{} has a smaller communication
cost than Graphene in all but the last group (which has the largest cardinality
of $\belamalis$),
by a factor of up to 7.4 (when $\diffsize = 10,000$).
This is because when $\diffsize$ is small, most of Graphene's communication
cost is used to transmit its IBLT, which, as our next experiment (on
bidirectional \problemname{}) also shows, is much larger than the sketch of
\solutionname{}.
When $\diffsize$ is large (at least $2.5\alissize$), \solutionname{}'s
communication cost is slightly larger than Graphene's.

\vspace{2pt}
\noindent\textbf{Bidirectional \problemname{}}\label{sssec:bidirectional-eval}
In this experiment, we set the universe size $|\universe|$ to $2^{256}$, which
is the same as the Ethereum dataset in~\autoref{ssec:ethereum}.
Among seven groups of experiments, $|\alisset|$ is fixed at one million,
$|\alismbela|$ is fixed at 10,000, whereas $|\belamalis|$ increases from 100 to
300,000.
For each group, $10,000$ instances of $\alisset$ and $\belaset$ are generated
randomly.
Bob (instead of Alice) initiates the \solutionname{} protocol (sending the
first message) in the first three groups (where $|\belamalis| < |\alismbela|$),
since as mentioned in~\autoref{ssec:ping-pong}, this leads to a smaller
communication cost than the other way around.

According to \Cref{fig:bidirect}, \solutionname{} consistently outperforms IBLT
and ECC by a factor of 7.8 and 4.7 respectively (at $|\belamalis|=10,000$) to
14.8 and 9.0 respectively (at $|\belamalis|=300,000$).
This, again, shows the low communication cost of \solutionname{}, as a
\problemname{} protocol, over \reconname{} protocols.
\solutionname{} uses 7.0 (at $|\belamalis|=3000$) to 8.6 (at
$|\belamalis|=300,000$)
rounds of communications on average.

\subsection{Results on Ethereum Data}\label{ssec:ethereum}
This experiment evaluates \solutionname{} in a large-scale bidirectional
\problemname{} problem derived from a dataset of Ethereum accounts as follows.
In the most common Ethereum synchronization protocol called \emph{state heal},
the \emph{world state trie} is a 16-ary prefix tree whose integrity is
protected by cryptographic hashing (i.e., a Merkle Patrica
trie~\cite{Merkle_merkletree_1987}).
This trie stores the account states of all Ethereum users in the world and is
shared by all Ethereum nodes.
When an Ethereum node recovers from a temporary offline period, it needs to
catch up with the latest world state before it can make new transactions.
According to~\cite{yang-rateless-iblt}, the state-of-the-art technique to
synchronize this trie is by IBLT.

According to~\autoref{sssec:delta-sync}, the detection of deltas is often the
first step to synchronizing files on cloud storage services.
Although the Ethereum blockchain is a slightly different context, we can define
the deltas as the different account states between two Ethereum nodes, say
Alice and Bob, and evaluate the detection of deltas, which is a large-scale
bidirectional \problemname{} problem in essence, on the publicly available
Ethereum data.
In other words, this experiment focuses on a thus induced bidirectional
\problemname{} problem rather than actually synchronizing the world state trie.

Specifically, we regard the state of each account as a 3-tuple consisting of
the account number, balance, and nonce (number of previous transactions
initiated by this account).
Given Alice's (or Bob's) \emph{snapshot} of the world state trie, we convert
the states of all Ethereum accounts therein into a set of 256-bit strings by
hashing the 3-tuple of every account into a 256-bit \emph{signature} using the
\texttt{SHA-256} cryptographic hash function~\cite{sha2}.

In this experiment, we use three snapshots of the world state trie downloaded
on different dates.
The three sets $\aset$ converted from these snapshots are denoted by
$\alisset$, $\belaset$, and $\coirset$ (from the newest to the oldest) and
their statistics are summarized in \Cref{tab:ether-state}.
We conduct two groups of experiments, in which Alice always holds the latest
snapshot $\alisset$, whereas Bob holds $\belaset$ (one day of staleness) and
$\coirset$ (more than fifty days of staleness), respectively in two groups.
Note that the latter group is much more challenging as a \problemname{} problem
(instance) than the former, since $|\alisset \triangle \coirset|$ is more than
24 times as large as $|\alisset \triangle \belaset|$.


In each group, we run \solutionname{} (with Bob initiating the protocol) and
IBLT~\cite{Eppstein_WhatsDifference_2011} (as suggested
by~\cite{yang-rateless-iblt}) for five times each (getting the correct set
intersection in every run) and summarize their average communication costs in
\Cref{tab:ethereum}.
The results show conclusively that \solutionname{} has a smaller communication
cost than IBLT by about an order of magnitude in both groups (regardless of the
SDC value).
These results also demonstrate that the outperformance over IBLT (which we have
shown in~\autoref{sssec:bidirectional-eval}) scales all the way to
cardinalities in hundreds of millions.

\begin{table}[!t]
  \centering
  \begin{tabular}{c c c c c c c}
    \toprule
    $\aset=$             & Block ID                     & Date              &
    $|\aset|$            & $|\aset
    \setminus \alisset|$ & $|\alisset \setminus \aset|$ & $|\aset \triangle
      \alisset|$
    \\
    \midrule
    $\alisset$           & \texttt{22399992}            & May 03, 2025      &
    292,222,740          & --
                         & --                           & --
    \\
    $\belaset$           & \texttt{22392874}            & May 02, 2025      &
    291,992,904          & 340,292
                         & 570,128                      & 910,420
    \\
    $\coirset$           & \texttt{22020359}            & March 11, 2025    &
    280,973,256          & 5,636,348
                         & 16,885,832                   & 22,522,180
    \\
    \bottomrule\end{tabular}
  \caption{Summary of the Ethereum dataset. 
  }\label{tab:ether-state}
\end{table}

\begin{table}[!t]
  \centering
  \begin{tabular}{l c c | l cc }
    \toprule
    Comm.
    Cost       & IBLT           & \solutionname{} & Comm.
    Cost       & IBLT           & \solutionname{}                                    \\ \midrule $\mathrm{\problemname}(\alisset,
    \belaset)$ & \SI{48.59}{MB} & \SI{5.866}{MB}  & $\mathrm{\problemname}(\alisset,
    \coirset)$ & \SI{1.192}{GB} & \SI{118.1}{MB}                                     \\ \bottomrule\end{tabular}
  \caption{Communication costs for \problemname{} on the Ethereum dataset,
    averaged from five runs of each protocol.
    IBLT uses two rounds of communications in all runs, whereas \solutionname{}
    always uses five.
  }\label{tab:ethereum}
\end{table}

As mentioned in the last paragraph of \autoref{sec:psea}, \solutionname{} pays a modest price in memory and running time 
in exchange for a much smaller communication cost than IBLT.
For example, the total running times on the two groups in \Cref{tab:ethereum} are roughly 100 and 165 
minutes respectively for \solutionname{}, compared to roughly 40 and 48.5 minutes respectively for IBLT.
If shorter running times are necessary, we can speed up \solutionname{} (and IBLT) by first partitioning the universe using a hash function like in PBS~\cite{gong-pbs}, 
and then computing the set intersections in all partitions in parallel (say using multiple cores).  The parallelization gain should grow linearly with the number of cores 
used, since there is no data dependency among the computation tasks;  and the increase in communication cost due to this partitioning should be tiny, as shown in 
the PBS case~\cite{gong-pbs}.
\section{Related Work}\label{sec:related}


In this section, we describe existing protocols on \problemname{}
(\autoref{ssec:setx-related}) and the two aforementioned related problems,
namely set membership (\autoref{ssec:membership}) and set reconciliation
(\reconname{},~\autoref{ssec:related-recon}).
We also compare \solutionname{} with CS-Sketch~\cite{cs-sketch} in \autoref{ssec:cs-sketch}, a recent sketching scheme for full traffic measurement that is also based on an RIP-1 CS matrix.
Finally, we compare \problemname{} with a related cryptographic problem named
private set intersection (PSI) in~\autoref{ssec:private}.

\subsection{Approximate Set Membership Filters (SMF)}\label{ssec:membership}
In the \emph{set membership} problem, Alice wants to summarize $\alisset$ into
a small filter, so that Bob, by querying this filter, knows whether an
arbitrary element $x$ belongs to $\alisset$.
Different from \problemname{} and \reconname{}, a set membership filter (SMF)
cannot exploit shared elements between Alice and Bob.
It has been shown that the size of exact SMFs (that answer all queries
correctly) is lower bounded at
$\log_2\binom{|\universe|}{|\alisset|}$~\cite{bordnik-membership}, so in large
universes, all practical SMFs are approximate.

Bloom filters (BF)~\cite{Bloom_Space/TimeTradeoffs_1970} are arguably the
best-known approximate SMF.
The filter is an array of $\counternum$ bits (cells) that are initialized to
zeros.
To encode BF, for each element $a\in\alisset$, $a$ is hashed to $\fanout$ cells
(by $\fanout$ independent hash functions) in the array, and these bits are set
to ones.
Finally, given an element $x$, the test (on whether $x$ belongs to $\alisset$)
is positive if and only if all $\fanout$ cells that $x$ is hashed to are ones.
It is well-known that BF tests have zero false negatives but occasional false
positives at a rate of $2^{-(\ln 2)\cdot|\alisset|/\counternum}$.

Counting Bloom filters (CBFs)~\cite{fan-cbf} extend BFs to support deletions of
elements from a BF.
In a CBF, each cell is a counter instead of a bit.
When an element $a\in\alisset$ is added (or deleted) from the CBF, all counters
that $a$ is hashed to are incremented (or decremented).
The resulting CBF can be tested similarly as BFs, by treating all nonzero
counters as ones.
However, deleting a false positive element from the CBF can lead to false
negative test results even for elements that are actually in $\alisset$.

Compressed BFs~\cite{mitzen-cbf} are designed to reduce the communication cost
of BFs.
Its idea is to populate the BF with only $\fanout=1$ hash function and to
compress the array using an entropy coding scheme (for example~\cite{duda-ans})
before sending it to Bob.
By~\cite{mitzen-cbf}, compressed BFs are about 15\% smaller than traditional
ones (under the same false positive rate), at a small computation cost of the
entropy coding.

\subsection{Set Reconciliation (\reconname{}) Protocols}\label{ssec:related-recon}
We have formulated the \reconname{} problem in~\autoref{sec:intro}.
Most existing \reconname{} protocols are based on either invertible Bloom
lookup tables (IBLT)~\cite{Eppstein_WhatsDifference_2011} or
error-correction codes (ECC).

\noindent\textbf{Invertible Bloom Lookup Tables}\label{f}
In an IBLT, each element (from a set) is inserted into $\fanout$ cells indexed
by $\fanout$ independent hash functions.
Whereas each cell is a single bit in a BF, it has three fields and takes up
about $1.5\bitwidth$ bits (if the universe size is $2^\bitwidth$) in an IBLT.
Therefore, IBLTs are much more powerful than BFs: The set difference
$\alisset\triangle \belaset$ of sets $\alisset$ and $\belaset$ can be recovered
from the ``difference'' of their IBLTs using a ``peeling process'' similar to
the belief propagation algorithm~\cite{pearl-bp} widely used in low-complexity
decoding of erasure-correcting codes, such as Tornado
codes~\cite{luby1998tornado}.
For this decoding process to succeed with a high enough probability, IBLT-based
\reconname{} protocols, such as Difference Digest
(D.Digest)~\cite{Eppstein_WhatsDifference_2011}, have to use roughly
$1.36\diffsize$ cells.
This translates into a communication overhead of roughly
$2.04\bitwidth\diffsize$, or more than double the theoretical minimum for
\reconname{}.

\noindent
\textbf{Set Reconciliation Protocols Based on Error Correction Codes}
We use PinSketch~\cite{Dodis_PinSketch_2008} as an example to illustrate the
basic ideas of ECC-based
protocols~\cite{Minsky2003,Dodis_PinSketch_2008,Naumenko_ErlayEfficientTransaction_2019,Karpovsky_eccreconciliation_2003}.
Recall from~\autoref{ssec:special-overview} that $\alisvec$ and $\belavec$
denote the binary vector representations of Alice's set $\alisset$ and Bob's
set $\belaset$, respectively.
In PinSketch, Alice encodes $\alisvec$ into a BCH~\cite{bch1} codeword and
sends it to Bob.
Bob treats $\belavec$ as the ``corrupted message'' and finds out the positions
of bit flips between $\alisvec$ and $\belavec$ (which correspond to unique
elements in Alice or Bob) by decoding the BCH codeword.

ECC-based protocols use a smaller communication cost compared with IBLT-based
ones, but they have much higher computation complexity, mainly because of the
BCH decoder, whose best time complexity (so far) is
$O(\diffsize^2)$~\cite{Berlekamp2015}.
A recent protocol called PBS~\cite{gong-pbs} (which stands for parity bitmap
sketch) combines both protocols.
PBS partitions the universe $\universe$ with a hash function, uses an ECC-based
approach to identify partitions that contain unique elements, and reconciles
the unique elements within each such partition using an IBLT-like approach.
As a result, PBS uses slightly more communication than ECC-based protocols, but
has a low time complexity comparable to IBLT-based protocols.

\subsection{Existing \problemname{}
  Protocols}\label{ssec:setx-related} The following protocols were proposed for
solving ``set reconciliation,'' but in retrospect, they are in fact
\problemname{} protocols, since they compute the set intersection instead of
the union.

\noindent\textbf{Counting Bloom Filters}
The \problemname{} protocol proposed in~\cite{Guo_SetReconciliationvia_2013} is
based on counting Bloom filters (CBFs).
In this protocol, Alice sends $CBF(\alisset)$, a CBF that encodes $\alisset$,
to Bob, and Bob approximates $\belamalis$ with all elements in $\belaset$ that
pass the test of $CBF(\belaset) - CBF(\alisset)$ (the difference between Bob's
CBF that encodes $\belaset$ and the one received from Alice).
This protocol is related to \solutionname{}, because as mentioned
in~\autoref{ssec:hash-encoding}, when the CS matrix $\sensemat{}$ follows
\Cref{def:sense-mat}, Alice's sketch $\sensemat \alisvec$ in \solutionname{}
can be regarded as $CBF(\alisset)$, and furthermore, $\sensemat
  \indicator_\belamalis = \sensemat\belavec - \sensemat\alisvec$, from which Bob
reconstructs $\belamalis$ in \solutionname{}, can also be regarded as
$CBF(\belaset) - CBF(\alisset)$.

Although both protocols use the same sketches ($\sensemat \alisvec$ in
transmission and $\sensemat \indicator_\belamalis$ to compute $\belamalis$),
their recovery procedures and hence recovery qualities (of $\belamalis$) are
completely different.
In~\autoref{ssec:hash-encoding}, we have shown using the RIP-1 argument that
$\sensemat \indicator_\belamalis$ contains enough information for lossless
reconstruction of the exact $\belamalis$, and \solutionname{} achieves this
goal using the MP decoder in~\autoref{ssec:omp}.
The protocol in~\cite{Guo_SetReconciliationvia_2013}, however, treats
$\sensemat \indicator_\belamalis$ simply as a CBF (and hence its decoding has
nothing to do with CS), which does not make full use of the information
therein.
For this reason, it can only compute an approximate result that contains both
false positives and false negatives~\cite{fan-cbf}.
Moreover, \solutionname{} is technically much more general than the protocol
in~\cite{Guo_SetReconciliationvia_2013}, because it works also on other sparse
RIP-1 CS matrices (such as the one in~\cite{charikar-cs}, in which half of the
$1$'s are flipped to $-1$'s) whereas the latter works only for the ``CBF
matrix."

\noindent\textbf{Graphene}
Graphene~\cite{Ozisik2019}, the state-of-the-art protocol for unidirectional
\problemname{}, combines IBLT with BF.

Alice sends both a BF and an IBLT of $\alisset$ to Bob, with the latter encoded
in just enough number of cells for BF's false positives.
Bob first computes $\hat{\alisset}$, the set of elements in $\belaset$ that
pass the BF, deducts the elements of $\hat{\alisset}$ from the received IBLT,
and finally decodes the BF's false positives $\hat{\alisset} \setminus
  \alisset$ from the resulting IBLT.
Therefore, the set difference $\belamalis$ can be computed as the union of
$\belaset\setminus \hat{\alisset}$ and BF's false positives.

Graphene uses a small communication cost on a wide range of SDCs ($\diffsize =
  |\belamalis|$) for the following reason.
On the one hand, the BF's size is $O(|\alisset|)$, but it cannot compute the
exact set intersection because of its false positives.
On the other hand, the IBLT is exact, but its size is $O(\diffsize \bitwidth)$.
In most scenarios, $\diffsize$ is not too small, so the BF is smaller than the
IBLT.
By combining both protocols, Graphene can compute the exact set intersection
without paying a large communication cost.
Note that when $\diffsize$ is very small, the IBLT becomes smaller than the BF,
in which case Graphene drops the BF and degenerates to IBLT.
In comparison, \solutionname{} only uses
$O(\belasize\log(\belasize/\diffsize))$ bits of communication, which, as shown
in \Cref{fig:unidirect}, is much smaller than that used by Graphene unless
$\diffsize$ is very large.
Finally, we comment that, as mentioned in~\autoref{sec:psea}, Graphene actually
solves a unidirectional \problemname{} problem (of computing
$\hat{\alisset}\setminus \alisset$ with $\alisset \subseteq \hat{\alisset}$)
using IBLT as a subroutine, to which end \solutionname{} uses much less
communication cost than IBLT.

\subsection{Comparison with CS-Sketch}\label{ssec:cs-sketch}
CS-Sketch~\cite{cs-sketch} is a recent sketching scheme for full traffic measurement.
The goal of that problem is, given a stream of packets in the data plane of a network device, count the number of packets in each flow, which is identified by a unique 5-tuple (source IP, destination IP, source port, destination port, and protocol).

In~\cite{cs-sketch}, the set of all unique 5-tuples (flow labels) can be identified accurately using a Bloom filter.
The nontrivial problem is to record and recover the count associated with each flow label (signal).
In CS-Sketch, the recording scheme is the same as that of Count-Min sketch~\cite{countmin}, which is equivalent to updating a CS codeword when the CS matrix is a sparse binary one.
According to~\cite{cs-sketch}, their CS matrix can also be viewed as the adjacency matrix of a binary expander graph, so it also has the RIP-1 for a similar reason as shown by \Cref{th:rip1}.
The recovery (decoding) algorithm is based on OMP (orthogonal matching pursuit), and the authors propose to reduce the decoding time complexity by inverse Cholesky decomposition.

Although both \solutionname{} and CS-Sketch use a similar CS matrix and rely on the RIP-1 for accurate reconstruction, they are fundamentally different in the following two aspects.
First, the signal to recover (decode) is very different. In CS-Sketch, the support of the signal (indices of nonzero scalars) are already known, and the goal is to recover the flow sizes (values of these scalars).
In \solutionname{}, however, the scalars are zero-one, but the support corresponds to a very large unknown set ($\belamalis$ for Bob).
Second, their use very different algorithms to account for the differences in signal.
CS-Sketch uses OMP to accurately recover scalar values, whereas \solutionname{} uses MP adapted to binary signals (\Cref{def:us-update}) to efficiently identify the support of the signal.

\subsection{Private Set Intersection}\label{ssec:private}
Private set intersection (PSI) is a cryptographic problem that is loosely
related to \problemname{}.
In PSI, two hosts, say Alice and Bob, want to collectively compute the
intersection $\alisset \cap \belaset$ without revealing any information about
the \emph{private} (unique) elements ($\alismbela$ for Alice, and $\belamalis$
for Bob) to the other host~\cite{private-intersection}.
As a result of this privacy requirement, the design of PSI protocol is a
completely different research problem than \problemname{}.
According to a recent lecture note~\cite{orlandi-psi-lecture}, all existing PSI
protocol therein are based on public key cryptography and they use a much
larger communication cost than \solutionname{}.
\section{Conclusion}\label{sec:conclusion}

In this paper, we show, from both a theoretical and a practical perspective,
that \problemname{} (\problemnamelong{}) is fundamentally much cheaper than
\reconname{} (\reconnamelong{}), debunking a longstanding unspoken perception
that they are equally difficult in costs.
To this end, we develop a novel \problemname{} protocol called \solutionname{},
the communication cost of which handily beats the information-theoretic lower
bound of \reconname{}.
We also identify five example applications, in all of which \solutionname{}
improves over existing \reconname{}-based protocols.
In addition, we identify a CS matrix that is both friendly to applications and compliant to the
existing theories on CS.
Finally, we show, by extensive experiments on both synthetic and real life
datasets, that \solutionname{} has a smaller communication cost than the state of the
arts by a factor of up to 7.4 on unidirectional \problemname{} and around an
order of magnitude on bidirectional \problemname{}.

\section{Acknowledgment}
We thank our shepherd Prof. Hans van den Berg, and anonymous reviewers, for their insightful comments that have helped to improve the quality and the presentation of this work.  This work was supported in part by the National Science Foundation under Grant No. CNS-2007006 and by a seed gift from Dolby Laboratories.

\balance
\bibliographystyle{elsarticle-num}
\bibliography{bib/NTGfull,bib/set_difference,bib/added_during_revision}

\appendix
\newpage
\section{Comparison of Vanilla MP, SSMP, and Our Decoder with an Example}\label{sec:example}

In this section, we explain the following two claims that we made
in~\autoref{ssec:omp}.
The first claim is that, the several MP-like decoders proposed specifically for RIP-1
matrices~\cite{indyk-emp, berinde-smp, berinde-ssmp} are slower than the
vanilla MP decoder (\autoref{algo:omp}).
This is because they use $L_1$-pursuit (to be elaborated shortly), which is
more expensive computationally than the $L_2$-pursuit used by the vanilla MP
decoder.
The second claim is that, by restricting to binary signals, the vanilla MP decoder becomes almost
as capable as these specific ($L_1$-pursuit-based) decoders when the CS matrix
is sparse (having the RIP-1).
We will illustrate this fact using an intuitive example.

Recall from~\autoref{ssec:omp} that in each iteration, the vanilla MP decoder
selects $i^*$ and $\delta^*$ that minimizes the $L_2$ residue error $\|\cscode
  - \delta\sensecol_i\|_2$.
We refer to this design as \emph{$L_2$-pursuit}.
In contrast, in these RIP-1-specific decoders, the ``loss function'' to be
minimized is the $L_1$ residue error $\|\cscode - \delta\sensecol_i\|_1$, so we
refer to this design as \emph{$L_1$-pursuit}.
This difference in the norm has an observable impact on the running time,
because in $L_2$-pursuit, the optimal pursuit step $\delta^*$ is the
\emph{average} of several (determined by $i^*$) coordinates in
$\cscode$~\cite{tropp-greed-is-good}, whereas in $L_1$-pursuit, $\delta^*$ is
the \emph{median} of these coordinates~\cite{berinde-smp} (see \Cref{example}).
As we will show in\appendixref{sec:fast-omp}, to efficiently implement Bob's MP
decoder, the optimal pursuit step $\delta$ of all coordinates $i\in\belaset$
are stored in a priority queue.
For $L_2$-pursuit, whenever $\cscode$ is updated, every affected $\delta$
(average) can be updated in $O(1)$ time.
In contrast, for $L_1$-pursuit, every affected $\delta$ (median) needs to be
recomputed from scratch.


Now, we illustrate why $L_1$-pursuit and our design are more capable than
$L_2$-pursuit when the CS matrix is sparse using the following example.

\begin{example}\label{example}
  In this example, $\sensemat$ is a $7\times 3$ sparse binary matrix that has the
  RIP-1 but not the RIP.
  The measurement $\cscode_0$ on the LHS is encoded from the (ground-truth)
  binary signal $\signal_0 = \begin{pmatrix} 1 & 1 & 1\end{pmatrix}^T$.
  \begin{equation}\label{eq:example} \begin{pmatrix} 3 \\ 1 \\ 1 \\ 1 \\ 1 \\ 1 \\ 1
    \end{pmatrix} = \begin{pmatrix}
      1 & 1 & 1 \\
      1 & 0 & 0 \\
      1 & 0 & 0 \\
      0 & 1 & 0 \\
      0 & 1 & 0 \\
      0 & 0 & 1 \\
      0 & 0 & 1\end{pmatrix} \cdot \begin{pmatrix} 1
      \\ 1
      \\ 1\end{pmatrix} \quad(\cscode_0 = \sensemat \signal_0).
  \end{equation}

  \begin{itemize}
    \item By $L_1$-pursuit (SSMP), the three coordinates of $\signal$ result in the same
          $L_1$ residue error, so without loss of generality, we select $i^*=1$.
          Then, one can check that $\delta^*$ is the \emph{median} of the first three
          coordinates in $\cscode$, which is 1.
          Hence, the pursuit error on $x_1$ (the first coordinate of $\vec{x}$) is zero.

    \item By $L_2$-pursuit (vanilla MP), $i^*=1$ is selected similarly as above.
          The $\delta^*$ to minimize the $L_2$ residue error is the \emph{average} of the
          first three coordinates in $\cscode$, which is $5/3$.
          Hence, the pursuit error on $x_1$ is $|5/3-1| = 2/3$.

    \item By $L_2$-pursuit on binary signals (our decoder by \Cref{def:us-update}), $x_1$ will be updated to 1.
          The pursuit error in this case is also zero.
  \end{itemize}
\end{example}

As \Cref{example} shows, the key benefit of $L_1$-pursuit is that its
$\delta^*$ (as a median) is robust to ``noisy'' coordinates in $\cscode$ (which
is common when $\cscode$ is measured from an RIP-1 matrix), whereas that in
$L_2$-pursuit (as an average) is not.
As we mentioned in~\autoref{ssec:omp}, computing the correct $\delta^*$
robustly is crucial to the reconstruction capability.

When it comes to our design ($L_2$-pursuit on binary signals), it is also
robust to such noises (since zero-one signals are much more robust to noises
than analog signals), and more importantly, it inherits the high decoding speed
of $L_2$-pursuit.

\section{Proof of Decoding Time Complexity}\label{sec:fast-omp}
In this section, we analyze the computation complexity of our MP decoder.
We derive the following bound under an assumption that we have verified in all
of our experiments.

\begin{theorem}\label{th:decode-time}
  It takes the MP decoder in \solutionname{} (\autoref{algo:omp} combined with
  \Cref{def:us-update}) \\$O(\belasize(\log\belasize) \log(\belasize/\diffsize))$
    time to recover a signal, assuming that decoding completes in $O(\diffsize)$
  iterations.
\end{theorem}

We follow the proof of SSMP's time complexity in~\cite{berinde-ssmp}.
We focus on Bob's MP decoder, since by~\autoref{eq:assumption} ($\alissize \le
\belasize$), he has a larger set of signal coordinates to decode.
Before starting, we need to specify an important implementation design in our
decoder.

As suggested in~\cite{berinde-ssmp}, Bob uses a priority queue as follows to
avoid solving $L_2$-minimization in every iteration.
This priority queue stores the pair $(i, \delta_i)$ for each signal coordinate
$i\in \belaset$, wherein $\delta_i$ is the optimal pursuit step on the $i^{th}$
signal coordinate under the current residue $\cscode$, using the $L_2$ residue
error $\|\cscode - \delta_i \sensecol_i\|_2$ as the priority.
In each iteration, it suffices to pop the pair $(i^*, \delta_{i^*})$ with the
\emph{lowest} priority from the queue, update $\cscode$ according to
\Cref{def:us-update}, and to \emph{update all priorities} affected by the
changes in $\cscode$.
This priority queue can be implemented as a balanced binary search tree, in
which each insertion, deletion, or priority update takes $O(\log n)$ time ($n$
being the number of inserted items).

Actually, $\delta_i$ can also be used as the priority (in place of $\|\cscode -
  \delta_i \sensecol_i\|_2$).
To see this, note the following fact by basic linear algebra (and also
$\|\sensecol_i\|_2^2=\fanout$ by \Cref{def:sense-mat}),
\begin{equation}\label{eq:ds-i} \delta_i \triangleq
  \arg\min_{\delta\in\mathbb{R}} \|\cscode-\delta\sensecol_i\|_2 =
  \frac{\cscode^T \sensecol_i}{\|\sensecol_i\|_2^2} = \frac{\cscode^T \sensecol_i
  }{\fanout}; \quad \|\cscode-\delta_i\sensecol_i\|_2 = \sqrt{\|\cscode\|_2^2 -
    m\delta_i^2}.
\end{equation}
As the latter equation in~\autoref{eq:ds-i} shows, under the current $\cscode$
(fixed for all signal coordinates $i\in\belaset$), the $L_2$ residue error
(priority) is a monotonically decreasing function of $\delta_i$.
Hence, it is equivalent to use $\delta_i$ as the priority instead of the $L_2$
residue error.

\begin{proof}[Proof of \Cref{th:decode-time}]
  At the very beginning of decoding, Bob needs to populate this priority queue,
  which involves computing $\delta_i=\cscode^T \sensecol_i/\fanout$ for every
  $i\in \belaset$ and inserting them to the queue.
  The time complexity of doing so is $O(\belasize \log\belasize)$.

  It remains to show that the time complexity of each iteration is
  $O(\belasize(\log\belasize) \log(\belasize/\diffsize)/ \diffsize)$, so the time
  complexity of running $O(\diffsize)$ iterations is as claimed in
  \Cref{th:decode-time}.
  As mentioned above, each iteration involves popping the pair $(i^*,
    \delta_{i^*})$ from the priority queue, which takes $O(\log \belasize)$ time.
  Then, it takes $O(\fanout) = O(\log(\belasize/\diffsize))$ time to update
  $\cscode$ by \Cref{def:us-update}.
  Finally, note that the number of signal coordinates $j$ with $\delta_j$
  affected by this update in $\cscode$ (i.e., those with
  $\sensecol_i^T\sensecol_{j}\neq 0$) is $O(\belasize \log(\belasize/\diffsize)/
    \diffsize)$, because for each of the $O(\belasize)$ signal coordinates $j\neq
    i$, the probability that $\sensecol_i^T\sensecol_{j}\neq 0$ is $1-
    \prod_{s=0}^{\fanout -1}(\counternum -\fanout-s) / (\counternum-s) =
    O(\fanout^2 / \counternum) = O(\log(\belasize/\diffsize) /\diffsize)$.
  For each such signal coordinate $j$, the time complexity to update $\delta_j$
  is $O(\log \belasize)$: We can list every such $j$ in $O(1)$ time per item
  using a reverse lookup table, compute the new value of $\delta_j$ in $O(1)$
  time, and then update the priority of $(j, \delta_j)$ in $O(\log \belasize)$
  time.
  As a result, the total time complexity to update all priorities is
  $O(\belasize(\log\belasize) \log(\belasize/\diffsize)/ \diffsize)$.
\end{proof}

Note that the decoding time complexity in \Cref{th:decode-time} dominates the
$O(\belasize \log(\belasize/\diffsize))$ time complexity for Bob to compute
$\sensemat \belavec$ and then $\cscode = \sensemat\belavec - \sensemat\alisvec$
as mentioned in~\autoref{ssec:hash-encoding}.
In bidirectional \solutionname{}, suppose that the ping-pong decoding uses $R$
rounds of communications in total;  then Bob's total decoding time complexity is
$O(\belasize (\log\belasize) (\log(\belasize/\diffsize)+R))$, since the
priority queue needs to be populated for $O(R)$ times (every time Alice sends a
residue to him).

In fact, if decoding does not complete after $O(\diffsize)$ iterations, we can
fall back to SSMP, whose time complexity is also $O(\belasize(\log\belasize)
  \log(\belasize/\diffsize))$ (but it is slower than our decoder in practice, as
explained in\appendixref{sec:example}).
For this reason, our decoding of $\sensemat \indicator_\belamalis$ in
unidirectional \solutionname{} is guaranteed to complete in
$O(\belasize(\log\belasize) \log(\belasize/\diffsize))$ time as we claimed in
\Cref{th:uni-main}.

Finally, we remark that we pledge to release the source codes (in C++) of this
fast implementation of MP decoder (based on priority queues), but in
compliance with the anonymity requirements, we are not able to do so when this
draft is under review.

\section{Efficient Entropy Coding (Compression) of Messages}\label{sec:compress}
In the main body of this paper, we have only described the messages transmitted in
\solutionname{} (as shown in \Cref{fig:workflow}), assuming that they can be
compressed to their entropy values using an \emph{entropy coding} technique.
In this appendix, we elaborate on the entropy coding component in our
implementation of \solutionname{}.
In fact, almost all messages are compressed using a general technique named
asymmetric numerical systems (ANS,~\autoref{sssec:ans}), The only exception is
Alice's sketch $\sensemat \alisvec$, which requires a more sophisticated compression
scheme to be described in~\autoref{sssec:differential-coding}.

\subsection{Lossless Compression via Asymmetric Numeral
  Systems}\label{sssec:ans}
\emph{Entropy coding} (source coding) is a fundamental communication technique.
The message in transmission is usually regarded as a sequence of i.i.d.
symbols.
If both the sender and the receiver know the \emph{symbol distribution}, an
entropy coding technique allows the message to be losslessly transmitted in a
small communication cost that is only slightly larger than its entropy.
Among existing entropy coding techniques~\cite{huffman,rissa-arith-coding},
asymmetric numerical systems (ANS)~\cite{duda-ans} achieves the
state-of-the-art compression ratio and encoding/decoding time complexity.
Our implementation uses a fast ANS scheme named rANS~\cite{rans-tricks}, which
requires only one integer multiplication operation to encode or decode each
symbol.

Now, we describe how the residue $\cscode_{(t)}$ ($t=2,3,\ldots$) in
\Cref{fig:workflow} is compressed using entropy coding.
As we mentioned at the end of~\autoref{ssec:hash-encoding}, each coordinate of
$\cscode_{(t)}$ (symbol) is a discrete (integer-valued) random variable, which,
as we will show shortly, follows Skellam distribution~\cite{skellam}, but its
exact parameters are unknown to Alice and Bob.
The issue is that, all entropy coding schemes rely on these parameters in order
to achieve the maximum possible compression ratio.
To address this issue, we estimate these parameters by means of statistical
inference.
Our experiments show that using these estimated parameters has a negligible
impact on the final communication cost of \solutionname{}.

According to~\autoref{eq:bob-noise}, $\cscode_{(t)}$ can be written as
$\sensemat \indicator_{\mathsf{P}} - \sensemat\indicator_{\mathsf{N}}$, wherein
$\mathsf{P}$ (means the ``positive'' signal component) consists of Bob's
hallucinations and Alice's false negatives; and $\mathsf{N}$ (``negative''
component) consists of Bob's false negatives and Alice's hallucinations.
Since $\mathsf{P}$ and $\mathsf{N}$ are \emph{disjoint}, with $\sensemat$ by
\Cref{def:sense-mat}, the \emph{symbol distribution} $X$ is approximately
$\mathrm{Skellam}(\mu_1,\mu_2)$ (which is equal to $\mathrm{Poisson}(\mu_1)
  -\mathrm{Poisson}(\mu_2)$~\cite{skellam}) wherein $\mu_1 = |\mathsf{P}|
  \fanout/ \counternum$ and $\mu_2 = |\mathsf{N}| \fanout/ \counternum$.

In practice, $\mathsf{P}$ and $\mathsf{N}$ (hence $\mu_1$ and $\mu_2$) are
unknown without knowledge of the ground truth, but we can accurately infer
$\mu_1$ and $\mu_2$ from $\cscode_{(t)}$ (by the sender) using the following
\emph{method of moments} estimator: $\hat{\mu_1} = \bar{X} + \tilde{S}_X^2$ and
$\hat{\mu_2}=\bar{X} - \tilde{S}_X^2$~\cite{casella-stt-inf}, wherein $\bar{X}$
and $\tilde{S}_X^2$ are respectively the sample mean and sample variance of the
coordinates in $\cscode_{(t)}$.

\subsection{Effective Compression of Alice's Sketch}\label{sssec:differential-coding}
Now, we focus on Alice's sketch $\sensemat \alisvec$, which is always the
longest message in \solutionname{} (and the only one in the unidirectional
case).
As a result, effectively compressing $\sensemat \alisvec$ to a small size is
crucial to reducing the total communication cost of \solutionname{}.
However, the general entropy coding scheme above is ineffective on this task,
because it disregards the following \emph{common knowledge} between Alice and
Bob.
We consider a specific coordinate in Alice's measurement $\sensemat \alisvec$
and denote this random variable by $X$ similarly as above.
The corresponding coordinate in Bob's measurement $\sensemat \belavec$, denoted
by $Y$, is actually \emph{strongly correlated} to $X$, since the SDC $\diffsize
  \ll |\abintersec|$ (by~\autoref{eq:assumption}).
Hence, according to information theory, the \emph{mutual information} between
$X$ and $Y$ is nonnegligible~\cite{gamal-inf-theory}.

In \solutionname{}, we make use of this mutual information to reduce the
communication cost, by combining the following two techniques.
The first technique, called \emph{statistical
  truncation}~\cite{Hua_simplerbetterdesign_2012}, relies on the fact that the
realization of $Y-X$, which follows $\mathrm{Skellam}(\mu_1,\mu_2)$ (with
$\mu_1 = |\belamalis| \fanout/ \counternum$ and $\mu_2 = |\alismbela| \fanout/
  \counternum$) as we just mentioned, falls into a small range (which we denote
by $[v, w]$) with high probability.
Its idea is as follows.
Alice truncates $X$ to its remainder $\tilde{X}$ divided by $w-v+1$.
Since the range $[v, w]$ is small, the entropy $H(\tilde{X})$ is much less than
$H(X)$.
As a result, Alice can effectively compress $\tilde{X}$ using entropy coding
and then send it to Bob.
Upon receiving $\tilde{X}$, Bob recovers $\hat{X}$ so that $v\le Y-\hat{X}\le
  w$ and $Y-\hat{X}$ is congruent to $Y-\tilde{X}$ modulo $w - v +1$ (such value
is always unique).
It can be proved, using basic mathematics, that $\hat{X} = X$ if and only if
$v\le Y-X \le w$.

Our second technique is an \emph{error correction code} (ECC, and specifically
we use the BCH code~\cite{bch1,bch-github} in our implementation) that patches
the errors between $X$ and $\hat{X}$ in the uncommon case of $Y-X \notin [v,
    w]$.
We describe our scheme of patching the parity (least significant) bit, and it
can be generalized to other (more significant) bits.
Note that when truncating $X$ to $\tilde{X}$, Alice also get a sequence of
integer quotients $X /(w-v+1)$ for each coordinate $X$ in $\sensemat\alisvec$.
She encodes the \emph{parities} of these quotients into an ECC codeword and
sends it to Bob along with $\tilde{X}$'s.
Bob can also compute a sequence of parities similarly as Alice (from the
quotients of $\sensemat \belavec$) and then recover Alice's parities by
correcting his own parities using the received ECC codeword.
Here, this ECC codeword uses much less communication cost than directly sending
the parity bits, similarly to the idea of
PinSketch~\cite{Dodis_PinSketch_2008}.
After performing statistical truncation, Bob checks whether each recovered
$\hat{X}$ agrees with Alice's parity of $X$.
If not, $\hat{X}$ is corrected to the \emph{most likely} value that satisfies
both the parity and congruence (to $Y-\tilde{X}$) requirements.

Finally, we remark that \solutionname{} can tolerate occasional errors (in
which $\hat{X}\neq X$) caused by decoding failures of the ECC.
Our MP decoder first treats these errors as noises.
If it cannot make further progress, but the residue is still not zero, it
corrects all coordinates of the residue to their (respective) congruent values
in $[v, w]$ (and maintains this invariant from this point on) and continues
with $L_1$-pursuit, since the median therein (see \Cref{example}) is robust to
occasional errors (even though may have very large absolute values).
This strategy is shown to be effective in our experiments.

\end{document}